\documentclass[preprint,showpacs,preprintnumbers,amsmath,amssymb,superscriptaddress,nofootinbib]{revtex4-1}
\usepackage{graphicx,color}
\usepackage{amsmath,amssymb,slashed}
\usepackage{url}
\usepackage{epstopdf}
\newcommand{\hs}{\hspace*{0.5cm}}

\newcommand{\be}{\begin{equation}}
\newcommand{\ee}{\end{equation}}
\newcommand{\bea}{\begin{eqnarray}}
\newcommand{\eea}{\end{eqnarray}}
\newcommand{\nn}{\nonumber}
\newcommand{\crn}{\nonumber \\}
\newcommand{\non}{\nonumber}

\newcommand{\la}{\lambda}

\newcommand{\fr}{\frac}

\newcommand{\bc}{\begin{center}}
\newcommand{\ec}{\end{center}}

\newcommand {\ba}{\begin{array}}
\newcommand {\ea}{\end{array}}
\newcommand{\ben}{\begin{enumerate}}
\newcommand{\een}{\end{enumerate}}

\usepackage{bm}
\usepackage{dcolumn}

\begin{document}

\title{ $(g-2)_{e,\mu}$ anomalies and decays  $h, Z\to e_b e_a $  in  3-3-1 models with  inverse seesaw neutrinos}

\author{T.T. Hong} \email{tthong@agu.edu.vn}
\author{L.T.T. Phuong} \email{lttphuong@agu.edu.vn}
\affiliation{An Giang University, Long Xuyen City 880000, Vietnam} 
\affiliation{Vietnam National University, Ho Chi Minh City 700000, Vietnam} 
\author{T.Phong Nguyen} \email{thanhphong@ctu.edu.vn}
\affiliation{Department of Physics, Can Tho University, 3/2 Street, Ninh Kieu, Can Tho City 94000, Vietnam}
\author{N.H.T. Nha} \email{nguyenhuathanhnha@vlu.edu.vn}
\author{L. T. Hue \footnote{Corresponding author}} \email{lethohue@vlu.edu.vn}
\affiliation{Subatomic Physics Research Group, Science and Technology Advanced Institute, Van Lang University, Ho Chi Minh City, Vietnam}
\affiliation{Faculty of Applied Technology, School of Technology,  Van Lang University, Ho Chi Minh City, Vietnam} 
\begin{abstract}
  The lepton flavor violating (LFV) decays $h, Z\to e_b e_a $, and $e_b\to e_a \gamma$ are discussed in a class of general  3-3-1 models adding heavy neutral leptons and singly charged Higgs bosons to accommodate experimental data of neutrino oscillation and $(g-2)_{e_a}$ anomalies of charged leptons through the inverse seesaw mechanism. We show that the models with the minimal number of inverse seesaw neutrinos cannot reach the $1\sigma$  range of $(g-2)_{\mu}$ data due to the experimental upper bounds on decay rates of  $(\tau \to \mu \gamma)$ and $(\mu \to e \gamma)$.  Features of LFV decays are presented in detail, focusing on the  regions of parameter space satisfying the $1\sigma$ ranges of  $(g-2)_{e,\mu}$ data. 
 
\end{abstract} 
\maketitle
 \section{\label{intro} Introduction}
 \allowdisplaybreaks
 
 Based on the original works \cite{Singer:1980sw, Pleitez:1992xh, Foot:1994ym, Ozer:1995xi}, a  general class  of   3-3-1 models constructed  from the gauge group $SU(3)_C\otimes SU(3)_L \otimes U(1)_X$ was  introduced \cite{Diaz:2003dk, Diaz:2004fs, Buras:2012dp}, in which  new exotic leptons carrying  arbitrary electric charges as functions of a parameter $\beta$ (331$\beta$).  Various phenomenologies  of this model class were discussed  \cite{CarcamoHernandez:2005ka, Buras:2012dp,  Salazar:2015gxa, Hue:2017lak,Hung:2019jue,  Hue:2021zyw, Suarez:2023ozu,   Cherchiglia:2022zfy}, indicating  a large lower bound of $\mathcal{O}(10)$ TeV for $SU(3)_L$ scale and heavy singly charged Higss boson masses. Moreover, this 331$\beta$ model consisting of six new inverse seesaw (ISS) neutrinos, named the 331$\beta$ISS model,  and a singly charged Higgs boson can explain successfully both the $(g-2)_{e,\mu}$ data and the neutrino oscillation data.     In this work, we continue discussing properties of  the lepton flavor violating (LFV) decays of the standard-model-like (SM-like) and $Z$ bosons, denoted  as  LFV$h$ and LFV$Z$, respectively. The correlations with LFV decays of charged leptons (cLFV) $e_b\to e_a \gamma$ will be investigated also.  Other extensions of  3-3-1 models  with new vectorlike fermions, leptoquarks, or/and scalars \cite{Lindner:2016bgg, DeJesus:2020yqx, deJesus:2020ngn,  Hue:2021zyw, Doff:2024cap} can accommodate the $(g-2)_{\mu}$ anomalies and cLFV experimental constraints,  but  the correlations between LFV decays of $h$ and $Z$ bosons  and updated $(g-2)_{e_a}$ data have not been mentioned.    
 
Normally,  the 331ISS models  discussed previously consist of  six ISS neutrinos, although the minimal number  can be 4, well known as minimal ISS  (mISS)  models, which   reduce a significant number of independent couplings, therefore is predictive for the correlations between $(g-2)_{e,\mu}$ and LFV decays. The mISS models were  shown to  accommodate the neutrino oscillation data and the  observed baryon asymmetry from the presence of the minimal number of new neutral heavy  leptons with masses at TeV scale \cite{Chakraborty:2021azg, Chakraborty:2022pcc, Marciano:2024nwm,Gogoi:2023jzl}.  The LFV source will generate one-loop contributions to  cLFV decays such as $e_b\to e_a \gamma$, LFV$h$, LFV$Z$, etc \dots. In this work, we will introduce  LFV sources as well as LFV one-loop contributions  in the general 331$\beta$ISS$(K,K)$ framework  having $2K$ heavy ISS neutrinos. After that, we will focus on two simple frameworks of 331$\beta$mISS and the 331$\beta$ISS with $(3,3)$ ISS neutrinos, in which  the possibility to accommodate  both the $(g-2)_{e,\mu}$ data and LFV decay rates will be investigated numerically. The ISS mechanisms with $2K>6$ \cite{CentellesChulia:2020dfh} such as the double ISS one may be investigated in our future work. 
 
 Experimental data for $(g-2)_{e,\mu}$ anomalies   have been updated from Ref. \cite{Muong-2:2023cdq} showing a 5.1$\sigma$ standard deviation from the  SM prediction  \cite{Aoyama:2020ynm, Davier:2010nc, Danilkin:2016hnh,  Davier:2017zfy, Keshavarzi:2018mgv, Colangelo:2018mtw, Hoferichter:2019mqg, Davier:2019can, Keshavarzi:2019abf, Kurz:2014wya, Melnikov:2003xd, Masjuan:2017tvw, Colangelo:2017fiz, Hoferichter:2018kwz, Gerardin:2019vio, Bijnens:2019ghy, Colangelo:2019uex, Colangelo:2014qya, Blum:2019ugy, Aoyama:2012wk, Aoyama:2019ryr, Czarnecki:2002nt, Gnendiger:2013pva, Pauk:2014rta, Jegerlehner:2017gek, Knecht:2018sci,Eichmann:2019bqf, Roig:2019reh} that  $\Delta a^{\mathrm{NP}}_{\mu}\equiv  a^{\mathrm{exp}}_{\mu} -a^{\mathrm{SM}}_{\mu} =\left(2.49\pm 0.48 \right) \times 10^{-9}$ \cite{Venanzoni:2023mbe}.  The  experimental $a_e$ data  were  reported from different groups~\cite{Hanneke:2008tm, Parker:2018vye, Morel:2020dww, Fan:2022eto}, leading to the two inconsistent deviations  between experiments and the SM prediction \cite{Aoyama:2012wj,  Laporta:2017okg, Aoyama:2017uqe,  Terazawa:2018pdc, Volkov:2019phy, Gerardin:2020gpp}. In this work, we chose the value $\Delta a^{\mathrm{NP}}_{e} = \left( 3.4\pm 1.6\right) \times 10^{-13}$, using the latest experimental data  for  $ a^{\mathrm{exp}}_{e}$ \cite{Fan:2022eto}. Although another $a^{\mathrm{SM}}_{e}$ result was derived \cite{Parker:2018vye},  leading to a $3.9\sigma$ deviation that $\Delta a^{\mathrm{NP}}_{e} = \left(-10.2\pm 2.6\right) \times 10^{-13}$,  which  has  the same  magnitude order as the first result. Therefore, we choose one of them for illustration.

The cLFV rates are constrained from recent (future)  experiments as follows  \cite{BaBar:2009hkt, MEG:2016leq, Belle:2021ysv, MEGII:2023ltw} (\cite{MEGII:2018kmf, Belle-II:2018jsg}):  	$\mathrm{Br}(\mu\rightarrow e\gamma) < 3.1\times 10^{-13}$  ($6\times 10^{-14}$), $\mathrm{Br}(\tau\rightarrow e\gamma) <3.3\times 10^{-8}$ ($ 9 \times 10^{-9}$), and  $\mathrm{Br}(\tau\rightarrow \mu\gamma) <4.2\times 10^{-8}$ ($ 6.9 \times 10^{-9}$).  The latest experimental constraints for LFV$h$ decay rates are $	\mathrm{Br}(h\rightarrow \tau \mu) <1.5\times 10^{-3}$, $	\mathrm{Br}(h\rightarrow \tau e) <2\times 10^{-3}$, and  $\mathrm{Br}(h\rightarrow \mu e) <4.4\times 10^{-5}$ \cite{CMS:2021rsq, ATLAS:2019xlq, CMS:2023pte, ATLAS:2023mvd}. The future sensitivities at the high-luminosity Large Hadron Collider (HL-LHC) and $e^+e^-$ colliders may be  orders of  $\mathcal{O}(10^{-4}) $, $\mathcal{O}(10^{-4}) $, and $\mathcal{O}(10^{-5}) $  for the three above LFV$h$ decays, respectively \cite{Qin:2017aju, Barman:2022iwj, Aoki:2023wfb, Jueid:2023fgo}. 
 
  The latest experimental constraints for LFV$Z$ decay rates  are $	\mathrm{Br}(Z\rightarrow \tau^\pm \mu^\mp) <6.5\times 10^{-6} $,  $\mathrm{Br}(Z\rightarrow \tau^\pm e^\mp) <5.0\times 10^{-6} $, and $	\mathrm{Br}(Z\rightarrow \mu^\pm e^\mp) <2.62\times 10^{-7}$ \cite{ATLAS:2021bdj, ATLAS:2022uhq}.  
 The future sensitivities will be $10^{-6}$, $10^{-6}$, and $7\times 10^{-8}$ at HL-LHC   and $10^{-9}$, $10^{-9}$, and $10^{-10}$ at FCC-ee \cite{Dam:2018rfz, FCC:2018byv}, respectively. These LFV$Z$ decays were shown theoretically as promoting signals of new physics that are visible for incoming experimental searches  \cite{Korner:1992an,Ilakovac:2012sh, DeRomeri:2016gum,  Crivellin:2018mqz, Herrero:2018luu, Jurciukonis:2021izn,  Hernandez-Tome:2019lkb, Abada:2021zcm, Abada:2022asx, Hong:2023rhg, Calibbi:2021pyh}.  They will be discussed in the 331$\beta$ISS model, concentrating on the allowed regions of parameters successfully explaining  the $(g-2)_{e,\mu}$ data.

Our paper is organized as follows.  In Sec. \ref{sec:model} we  briefly review the 331$\beta$ISS$(K,K)$ model, determining all physical states, masses and mixing parameters of all leptons. In Sec. \ref{sec:Feynrules},  we  compute  all relevant couplings, Feynman rules  relating to  one-loop LFV contributions, and analytic formulas for $(g-2)_{e,\mu}$ anomalies and LFV decay rates. Interestingly, we introduce a general analytic expression  dominant one-loop contributions to $(g-2)_{e,\mu}$ and LFV amplitudes, arising from the well-known "chiral enhancement" ones. From this,  we will indicate a strict relation  between $\Delta a_{\mu}$ and Br$(\tau \to\mu \gamma )$ in  the 331$\beta$mISS framework. The numerical  illustrations for various relations among LFV decay rates and $\Delta a_{e,\mu}$ and important parameters of the two   331$\beta$mISS and 331$\beta$ISS are depicted in Sec. \ref{sec_numerical}. Important features discussed in our work are summarized in Sec. \ref{sec:con}.  Finally, there are four Appendices showing precisely the Higgs sector, master functions relating to formulas of $\Delta a_{e_a}$, and  general one-loop contributions to LFV$Z$ and  LFV$h$ decay amplitudes.  These formulas are constructed in the general case of $2K$ ISS neutrinos, being very useful for further studying the 331$\beta$ models consisting of various ISS neutrino kinds mentioned in Ref. \cite{CentellesChulia:2020dfh}.
 
\section{\label{sec:model} Review the 3-3-1 model with  ISS neutrinos}

Here, we summarize the most important content of the $331\beta$ISS$(K,K)$ model relating to our work. The lepton sector and charged lepton masses and mixing parameters were shown in detail in Ref. \cite{Hue:2018dqf}, where  left-handed (LH) leptons are assigned to antitriplets and right-handed (RH)  leptons are  singlets 
\begin{align}
&L'_{aL}= \begin{pmatrix}
	e'_a &  	-\nu'_{a}& 	E'_a 
\end{pmatrix}^T_L  \sim \left(3^*~, -\frac{1}{2}+\frac{\beta}{2\sqrt{3}}\right), \hs a=1,2,3,
\crn & e'_{aR}\sim   \left(1~, -1\right)  , \hs \nu_{IR},\; X_{IR}\sim  \left(~1~, 0\right) ,\hs E'_{aR} \sim   \left(~1~, -\frac{1}{2}+\frac{\sqrt{3}\beta}{2}\right),  \label{lep}
\end{align}
where we just pay attention to the $SU(3)_L\otimes U(1)_X$ group.  In general, we consider that the model includes  $2K$ RH neutrinos as $SU(3)_L$ singlets, namely,  $\nu_{IR}$ and $X_{IR}$ with $I=\overline{1,K}$. The prime denotes flavor states which are generally different from the  mass eigenstates. 

The quark sector is ignored here; see detail in Refs. \cite{Buras:2012dp, Hung:2019jue, Cherchiglia:2022zfy} for example.  The class of models constructed using $SU(3)_L$ antitriplet LH fermions   defined in Eq.~\eqref{lep} is  equivalent to the one using  triplets  through a transformation keeping physical results unchanged \cite{Descotes-Genon:2017ptp, Hue:2018dqf}.

The covariant derivative defined by the electroweak (EW) gauge group $SU(3)_L\otimes U(1)_X$ is  $D_{\mu}\equiv \partial_{\mu}-i g T^a W^a_{\mu}-i g_X X T^9X_{\mu},$ where $T^9=1/\sqrt{6}$, $g$, and $g_X$ are gauge couplings of  the two groups $SU(3)_L$ and $U(1)_X$, respectively. In general, the model 331$\beta$ model predicts  three
   charged gauge bosons: $ W^{\pm}_{\mu}=\frac{1}{\sqrt{2}}\left( W^1_{\mu}\mp i W^2_{\mu}\right)$, $Y^{\pm A}_{\mu}=\frac{1}{\sqrt{2}}\left( W^4_{\mu}\mp i W^5_{\mu}\right)$, and $V^{\pm B}_{\mu}=\frac{1}{\sqrt{2}}\left( W^6_{\mu}\mp i W^7_{\mu}\right)$. The new electric charges of the gauge and  Higgs bosons,  and charged leptons are $A=\fr{1}{2}+\fr{\sqrt{3}\beta}{2}$ and $	B=-\fr{1}{2}+\fr{\sqrt{3}\beta}{2}$. 
Three scalar triplets needed for generating masses are 
\begin{align}
\label{higgsc}
&\chi= \begin{pmatrix}
		\chi_{A} &
	\chi_{B} &
	\chi^0 
\end{pmatrix}^T \sim \left(3~, \frac{\beta}{\sqrt{3}}\right), 
\;  \rho= \begin{pmatrix}
	\rho^+ & \rho^0 & \rho^{-B} 
\end{pmatrix}^T \sim \left(3~, \frac{1}{2}-\frac{\beta}{2\sqrt{3}}\right),
\crn & \eta= \begin{pmatrix}
	\eta^0 & \eta^-& \eta^{-A}
\end{pmatrix}^T \sim \left(3~, -\frac{1}{2}-\frac{\beta}{2\sqrt{3}}\right), 
\end{align}
and a  new singly charged Higgs boson $ h^+\sim (1,1,1),$ needed for generating  large one-loop contributions  to $(g-2)_{e_a}$ anomalies.   Nonzero vacuum expectation values (VEVs)  of neutral Higgs components are   
\begin{align}
	\langle  \chi^0\rangle = \frac{v_3}{\sqrt{2}},\; \langle  \rho^0 \rangle= \frac{v_2}{\sqrt{2}}, \;  \langle   \eta^0 \rangle= \frac{v_1}{\sqrt{2}}.  \label{vevhigg1}
\end{align}
 In addition, two Higgs triplets can play roles as those in the well-known two-Higgs-doublet models (2HDM)  in generating masses for SM fermions in the 3-3-1 models \cite{Cherchiglia:2022zfy}.  Therefore, we use a notation appearing in the 2HDM that $	t_\beta\equiv \frac{v_2}{v_1}$, so that   $s^2_{\beta}+c^2_{\beta}=1$, following the notations  that $s_{x}\equiv \sin \theta_x$, $c_{x}\equiv \cos \theta_x$, and $t_{x}\equiv \tan \theta_x$, which will be used from now on. 

The symmetry breaking  patterns for the EW of the 331$\beta$ model are :  
$SU(3)_L\otimes U(1)_X\xrightarrow{v_3} SU(2)_L\otimes U(1)_Y\xrightarrow{v_1,v_2} U(1)_Q$, leading to  
the condition that  $v_3\gg v_1,v_2$. The first breaking step result in four massless gauge bosons  identified  with the SM ones. Consequently,  the $SU(3)_L$  gauge couplings are determined as 
\bea
t_X\equiv \fr{g_X}{g} = \fr{\sqrt{6}s_W}{\sqrt{1-(1+\beta^2)s_W^2}}, \; g=g_2,
\label{eq_beta_coupl}
\eea
where the weak mixing angle is defined as $t_W = \tan\theta_W = g_1/g_2$ and $g_{1,2}$ are the $U(1)_Y$ and $SU(2)_L$ gauge couplings of  the SM, respectively.
The experimental value of $s_W$ results in a constraint that $|\beta| \le \sqrt{3}$.
The masses of the charged gauge bosons are
\bea 
m^2_{Y} = \frac{g^2}{4}(v_3^2+v^{2}_1),\quad 
m^2_{V}=\frac{g^2}{4}(v_3^2+v^2_2),\quad
m^2_{W} = \fr{g^2}{4}(v^2_1 + v^{2}_2),
\label{masga}\eea 
where the gauge boson $W^\pm$ is identified with the SM one, implying that
\begin{equation}\label{eq_SMWmass}
	v \equiv \sqrt{v_1^2+v_2^2} =\frac{2 m_W}{g} \simeq 246\;\mathrm{ GeV}.
\end{equation}

The above Higgs sector is  enough to generate all SM quark masses and heavy new quark masses \cite{Buras:2012dp, Hung:2019jue, Cherchiglia:2022zfy}.  Recall here that  the Yukawa term $ Y^u_{3a}\overline{Q_{3L}}\rho^*u_{aR}$ generating the  tree-level  mass of top quark $m_t\simeq \frac{Y^u_{33}v_2}{\sqrt{2}} $,  implying a lower bound $t_\beta > 0.3$ corresponding to  the   pertubartive limit of $Y^u_{33}$:   $s_{\beta}\ge \sqrt{2}m_t/(\sqrt{4\pi}v) $, which is similar  to the 2HDM type II \cite{Aoki:2009ha}. 

The The Higgs sector of the 331ISS$(K,K)$ model is  summarized in Appendix \ref{app_Higgs}, consisting of  the singly charged Higgs part  \cite{Hue:2021zyw}, other heavy charged Higgs used from notations of Refs. \cite{Buras:2012dp, Hue:2017lak}.   The neutral Higgs sector  consists of the  SM-like Higgs boson inheriting properties consistent with experiments. 

The model predicts three neutral gauge bosons including the massless photon.  
The relation between the original and physical bases  of the  neutral gauge bosons is 
\begin{align}
	\begin{pmatrix}
		X_\mu \\
		W^3_\mu\\
		W^8_\mu
	\end{pmatrix} &= \begin{pmatrix}
		s_{331}c_W, &   \left(- s_{331}s_Wc_\theta+c_{331}s_\theta\right),  & \left( s_{331}s_Ws_\theta +c_{331}c_\theta\right) \\
		s_W ,& c_Wc_\theta ,& -s_\theta c_w \\
		c_{331}c_W, &   -\left( c_{331}s_Wc_\theta+s_{331}s_\theta\right),  &\left(  c_{331}s_Ws_\theta -s_{331}c_\theta\right) 
	\end{pmatrix}\begin{pmatrix}
		A_\mu \\
		Z_{1\mu} \\
		Z_{2\mu}
	\end{pmatrix}, 
	\label{neutralgaugebosonmix}
\end{align}
where $	s_{331}\equiv \sin\theta_{331}=\sqrt{1-\beta^2t_W^2}$, $c_{331}\equiv \cos\theta_{331} =\beta t_W $, and $t_X= \sqrt{6}t_W/s_{331}$.  In the limit $v^2/ v_3^2\to 0$,  the $Z-Z'$ mixing angle $\theta$ is  \cite{Buras:2014yna} 
\begin{align} \label{eq_stheta}
	s_{\theta}\equiv \sin \theta &= \left(3\beta t^2_W   +\frac{\sqrt{3}(t^2_{\beta}-1)}{t^2_{\beta} +1}\right)  \frac{\sqrt{1-\beta^2 t_W^2}v^2}{ 4 c_Wv_3^2},   
\end{align}
and  $M^2_{Z'}=g^2v^2_3/(3s^2_{331}) +\mathcal{O}(v^2)$.  The lower $SU(3)_L$ scale $v_3$ of  the  331$\beta$ model  was indicated to be as large as the order  $\mathcal{O}(10)$ TeV \cite{Okada:2016whh, deJesus:2023lvn, Long:2018fud, Alves:2022hcp, Suarez:2023ozu, CarcamoHernandez:2022fvl}. Hence,  the  neutral gauge bosons will be identified approximately as  $Z_2\equiv Z'$ and $Z_1\equiv Z$ found experimentally.

The Yukawa Lagrangian is \cite{Hue:2021zyw} 
\begin{align}
	\label{eq_ylepton1}
	-\mathcal{L}^{\mathrm{yuk}}_{\mathrm{lepton}} =&  \overline{e'_{R}} Y^e \eta^TL'_{L}  +  \overline{E'_{R}}Y^E \chi^T L'_{L} +  \overline{\nu_{R}}Y^{\nu} \rho^TL'_{L} 
\crn &+ \overline{\nu_R}M_R(X_{R})^c +\frac{1}{2} \overline{X_R}\mu_X(X_{R})^c  + \overline{(X_{R})^c} Y^{h}e'_{R} h^+ +\mathrm{h.c.}, 	
\end{align}
 The perturbative limit of the  matrix $Y^{h}$ requires that $|Y^{h}_{Ia}|<\sqrt{4\pi}\;\forall a=1,2,3$ and $I=\overline{1,K}$. The symmetric matrix $\mu_X$ is always chosen in the diagonal form $\mu_X=\mathrm{diag}(\mu_1,\mu_2,\dots, \mu_K)$ because of redefining the basis of $X_R$ and $M_R$.  We note that the general form of $M_R$  can be diagonalized  for determining the approximate  form of the total neutrino mixing matrix \cite{Hong:2022xjg}. In this work, $M_R$ is transformed into the symmetric form, using a rotation redefining the two bases $\nu_R$ and $X_R$, diagonalized generally through the following unitary transformations:
 \begin{align}
 \label{eq:newbase}
  \mu_X&\to U_X^T\mu_XU_X =\mathrm{diag}(\mu_1,\mu_2,\dots, \mu_K),\; X_R\to X_R=U_XX_R,
 \crn M_R&\to M_R=M_RU_X^\dagger \to M_R=M_R^T=U_RM_R,\;
 \nu_R \to \nu_R=U_R\nu_R.
 \end{align}

 Therefore,  we will work with assumptions that  $\mu_X$ is diagonal  and  $M_R$ is symmetric, $M_R=M_R^T$, leading to the unitary transformation $U_N$ into the diagonal form $\hat{M}_R$:
 \begin{equation}\label{eq:hatMR}
 U_N^TM_RU_N=\hat{M}_R= \mathrm{diag}(M_1,M_2,\dots,M_K). 
 \end{equation}
 
 In the  Lagrangian \eqref{eq_ylepton1},  the neutrino Dirac mass matrix comes from the third term including the Higgs triplet $\rho$ which also generates the top quark mass. The Dirac mass term and  the Yukawa couplings $Y^h$   depend on $t_{\beta}$  differently from those  mentioned in Ref. \cite{Hue:2021xap}. The lepton mass terms are 
\begin{align}
	-\mathcal{L}^\text{mass}_{\mathrm{lepton}}&=  \frac{Y^e_{ab}v_1}{\sqrt{2}} \overline{e'_{aR}}e'_{bL} +  \frac{Y^E_{ab}v_3}{\sqrt{2}} \overline{E'_{aR}} E'_{bL}
	 + \frac{1}{2} \begin{pmatrix}
		\overline{(\nu'_L)^c}& \overline{\nu_R} & \overline{X_R} 
	\end{pmatrix}
	\mathcal{M}^{\nu} 
	\begin{pmatrix}
		\nu'_L		\\
		(\nu_R)^c \\	
		(X_R)^c 	
	\end{pmatrix}
	+\mathrm{h.c.}, 
\crn  	\mathcal{M}^{\nu}  &= \begin{pmatrix}
		\mathcal{O}_{3\times 3}& m^T_D & \mathcal{O}_{3\times K}\\
		m_D&  \mathcal{O}_{K\times K}& M_R^T\\
		\mathcal{O}_{K \times3}&M_R& \mu_X
	\end{pmatrix}
	\; ,\label{eq_mlepton1}
\end{align}
where $ (m_D)_{Ib}\equiv  m_{D,Ib}= -\frac{Y^{\nu}_{Ib} v_2}{\sqrt{2}}$ for all $I=\overline{1,K}$, and 
 $$\nu'_L=(\nu'_1, \nu'_2, \nu'_3)_L^T, \; \nu_R=(\nu_1,\nu_2, \dots ,\nu_K)_R^T, \; X_R=(X_1,X_2, \dots ,X_K)_R^T.$$  The total mixing matrix is defined as 
\begin{align}
	\label{eq_Unu}
	U^{\nu T} 	\mathcal{M}^{\nu} U^{\nu } &= \hat{\mathcal{M}}^{\nu}=\mathrm{diag}(m_{n_1},\;m_{n_2},\;...,m_{n_{2K+3}}),
	\crn 	\begin{pmatrix}
		\nu'_L		\\
		(\nu_R)^c \\	
		(X_R)^c 	
	\end{pmatrix}  &= U^{\nu} n_{L},\; 
	\begin{pmatrix}
	\left(\nu'_L\right)^c		\\
	\nu_R \\	
	X_R 	
\end{pmatrix} = U^{\nu*} n_{R}= U^{\nu*} (n_{L})^c,\; 
\end{align} 
where $ n_{L,R}=(n_{1},n_{2},..., n_{2K+3})_{L,R}$ are Majorana neutrino mass eigenstates  $n_{iL,R}=(n_{iR,L})^c$.

In this work, we will consider the general case in which the flavor charged leptons are not physical, i.e.   left and right   $3\times3$ unitary  matrices $U^{e}_{L(R)}$ and $V^E_{L(R)}$ used to  change into the mass eigenstates of  SM and heavy leptons are defined as follows:
\begin{align}
e'_{L(R)}&=U^e_{L(R)}e_{L(R)}, \; 
%
 U^{e\dagger}_RY^eU^e_L = \frac{\sqrt{2}}{v_1} \hat{m}_{\ell} =\frac{\sqrt{2}}{v_1} \mathrm{diag}\left(m_e,\; m_{\mu},\; m_{\tau} \right) \label{eq:UeLR}
 \\ E'_{aL(R)} =& \left( V^{E}_{L(R)} \right)_{ab} E_{bL(R)}, \; V^{E\dagger}_RY^EV^E_L = \frac{\sqrt{2}}{v_3} \hat{m}_{E} =\frac{\sqrt{2}}{v_3} \mathrm{diag}\left(m_{E_1},\; m_{E_2},\; m_{E_3} \right). 
 \label{eq_Emass}
\end{align}
The total neutrino mixing matrix is parameterized in the  seesaw form
\begin{equation} 
	U^{\nu}= \left(
	\begin{array}{cc}
		\left(	I_3-\frac{1}{2}RR^{\dagger} \right) U^{\nu}_{3} & RV_N \\
		-R^\dagger U^{\nu}_3 & \left(I_{2K} -\frac{1}{2}R^{\dagger} R\right)V_N \\
	\end{array}
	\right)  +\mathcal{O}(R^3), 
	\label{eq_Unu0}	
\end{equation}
where    $V_N$ is a $2K\times 2K$ unitary matrix;   $R$ is a $3\times 2K$ matrix satisfying $|R_{aI}|<1$ for all $a=1,2,3$, and $I=1,2,\dots,2K$. The $3\times 3$ unitary matrix  $U_{\mathrm{PMNS}}$ is the Pontecorvo-Maki-Nakagawa-Sakata (PMNS) matrix, 
  defined as 
\begin{equation}\label{eq:Upmns}
U_{\mathrm{PMNS}}\equiv U^{e\dagger}_L  U^{\nu}_3 \to U^{\nu}_3=U^{e}_LU_{\mathrm{PMNS}}.
\end{equation}
Now, the Dirac and Majorana mass matrices  have   well-known ISS forms  as   \cite{Arganda:2014dta, Ibarra:2010xw}
\begin{align}
	M^T_D= (m^T_D,\hs \mathcal{O}_{3\times  K}), \hs M_N=\left(
	\begin{array}{cc}
		\mathcal{O}_{K\times K}& M_R \\
		M^T_R & \mu_X \\
	\end{array}
	\right),
	\label{repara}
\end{align}
and both $M_R$ and $\mu_X$ are $K \times K$ matrices.  The condition  max$|R_{ij}|\equiv ||R||\ll1$ gives the ISS relations approximately as follows:
\begin{align}
	&R= \left(\mathcal{O}_{3\times K},\hs m^\dagger_D\left(M^\dagger_R\right)^{-1} \right), \label{eq:R}
	\\ &\hat{m}_{\nu}= U^{\nu T}_3 m_{\nu}U^{\nu}_3 =  U^{\nu T}_3 m_D^T \left( M_R^T\right)^{-1}\mu_XM_R^{-1}m_DU^{\nu}_3 , \label{eq:mnu}
	\\& V^*_N\hat{M}_NV^{\dagger}_N  \simeq M_{N}.  
	\label{eq_RISS}
\end{align}
A general parametrization  of $m_D$ is based on  Eq. \eqref{eq:mnu} \cite{Casas:2001sr}, using  a   $K\times 3$ matrix $\xi$ defined as   $\xi \equiv \mu_X^{ \frac{1}{2}}M_R^{-1}m_DU^{\nu}_3\hat{m}_{\nu}^{-\frac{1}{2}}$, which obeys the orthogonal relation  $\xi^T\xi=I_3$ when  all  active neutrinos are massive. On the other hand, in the  mISS framework consisting of a massless neutrino, the formula  of $\hat{m}_{\nu}^{-\frac{1}{2}}$ needs to be  defined as 
\begin{align}
\label{eq:xi}
 \;\quad  \hat{m}_{\nu}^{-\frac{1}{2}}\equiv \left[ \begin{array}{c} 
\mathrm{diag}\left(0, m_{n_2}^{-\frac{1}{2}},\; m_{n_3}^{-\frac{1}{2}}\right)	\; (\mathrm{NO}) \\
\mathrm{diag}\left(  m_{n_1}^{-\frac{1}{2}},\; m_{n_2}^{-\frac{1}{2}},\;0\right)	\; (\mathrm{IO})	
\end{array} \right.,
\end{align} 
which satisfies  $\xi^T\xi=\mathrm{diag}(0,1,1)$ (NO) or $\xi^T\xi=\mathrm{diag}(1,1,0)$ (IO).  The NO and IO schemes are defined as two different ways of arranging the active neutrino masses that $m_{n_1}<m_{n_2}<m_{n_3}$ and $m_{n_3}<m_{n_1}<m_{n_2}$, respectively.
 
  The  analytic form of $m_D$ is derived generally as follows 
\begin{align}
\label{eq:mD22}
m_D&= x_0^{\frac{1}{2}}  M_R \hat{\mu}_X^{-\frac{1}{2}}\xi\hat{x}^{\frac{1}{2}}_{\nu}U^{\nu \dagger}_3,
\crn R&\simeq \left( \mathcal{O}_{3\times K},\; R_{0}\right), \; R_{0} \equiv x_0^{\frac{1}{2}} U^{\nu }_3 \hat{x}^{\frac{1}{2}}_{\nu}\xi^{\dagger}\hat{\mu}_X^{-\frac{1}{2}}, 
\end{align}
where $x_0$ is defined as the ratio of the two scales of $\mu_X$ and active neutrino masses, namely:
\begin{align}
\label{eq:x0}
x_0 &\equiv \frac{\mathrm{max}[m_{n_1},m_{n_2},m_{n_3}]}{\mathrm{max}[|(\mu_X)_{11,22}|]}\ll 1, 
\crn  \hat{\mu}_X & \equiv \frac{\mu_X}{\mathrm{max}[|(\mu_X)_{11,22}|]},\; \hat{x}_{\nu} \equiv \frac{ \hat{m}_{\nu}}{\mathrm{max}[m_{n_1},m_{n_2},m_{n_3}]} . 
\end{align}
The ISS condition $|\hat{m}_{\nu}|\ll |\mu_X|\ll |m_D|\ll M_{1,2}$  gives $\frac{\sqrt{\mu_X \hat{m}_{\nu}}}{M_{1,2}}\simeq0$ and $x_0\ll1$ but nonzero, which can be considered as the nonunitary scale of the active neutrino mixing matrix.

In the specific mISS framework  consisting of one active  massless neutrino,  the matrix $\xi$ with all real entries is shown to have  only the  form
\begin{align}
\label{eq:ximatrix}
\xi(\mathrm{NO})= & \begin{pmatrix}
0	& c_{\phi} & -s_{\phi} \\
0	&  s_{\phi}& c_{\phi}
\end{pmatrix}, \;  \xi(\mathrm{IO})=  \begin{pmatrix}
c_{\phi} & -s_{\phi} &0\\
 s_{\phi}& c_{\phi}&0
\end{pmatrix}, 
\end{align} 
where $c_{\phi} \equiv \cos \phi$ and $s_{\phi} \equiv \sin \phi$.

Based on Eqs. \eqref{eq:hatMR} and \eqref{eq_RISS},  we derive that   
\begin{align}
	\hat{M}_N &= \left(\begin{matrix}
		\hat{M}_R	& \mathcal{O}_{K\times K} \\ 
		\mathcal{O}_{K \times K}	& \hat{M}_R
	\end{matrix} \right), \;  	
%
  V_N \simeq \dfrac{1}{\sqrt{2}}
	\left(\begin{matrix}
		-iU_N 	& U_N  \\ 
		iU_N 	& U_N 
	\end{matrix} \right). \label{eq_UNmiss}	
\end{align}
Consequently, the approximate formula of $U^{\nu}$ ignoring the $ \mathcal{O}\left(|R_0|^3\right)$ part is  
\begin{align}
\label{eq:Usimeq}
U^{\nu}&\simeq  \begin{pmatrix}
\left(I_3 -\frac{R_0R_0^{\dagger}}{2}\right)U^{\nu}_3	& \frac{i R_0U_N }{\sqrt{2}}&  \frac{R_0U_N}{\sqrt{2}}\\
\mathcal{O}_{K\times3}	&  -\frac{iU_N}{\sqrt{2}}  &   \frac{U_N}{\sqrt{2}} \\
-R_0^{\dagger} U^{\nu}_3& 	\left(I_K -\frac{R_0^{\dagger}R_0}{2}\right) \frac{iU_N}{\sqrt{2}} & \left(I_K -\frac{R_0^{\dagger}R_0}{2}\right)\frac{U_N}{\sqrt{2}} 
\end{pmatrix} \equiv \begin{pmatrix}
U^{\nu}_0\\
U^{\nu}_1
\\
U^{\nu}_2
\end{pmatrix},
\end{align}
 where $U^{\nu}_k$ with $k=0,1,2$ used  for derivation  the Higgs couplings with leptons. 

The simplest case of $m_D$  is  $m_D=x_0^{\frac{1}{2}}  \hat{M}_R\hat{x}_\nu^{\frac{1}{2}}  U^\dagger_{\mathrm{PMNS}}$ with  $\mu_X=\mu_X I_K$,  $U^e_{L(R)}=I_3$,  and $M_R\equiv \hat{M}_R=M_0I_{K}$. Consequently, formulas of 
 $m_D$  and $R$ with the simplest form of $\xi$ are 
\begin{align}
	\label{eq_mDRISS}
	m_D&=x_0^{\frac{1}{2}} \hat{M}_R\hat{x}_\nu^{\frac{1}{2}}  U^\dagger_{\mathrm{PMNS}},
	\;  R \simeq \left(\mathcal{O}_{3\times K},\;  x_0^{\frac{1}{2}} U_{\mathrm{PMNS}}\hat{x}_\nu^{\frac{1}{2}}\right),
\end{align}
which is the most popular form  used to discuss on the LFV processes for standard ISS model with $K=3$. 
Using the ISS relation, the total neutrino mixing matrix given in Eq. \eqref{eq_Unu0}  are  determined approximately from $V$ and $R$ given in Eqs. \eqref{eq_UNmiss} and \eqref{eq_mDRISS}, respectively.  We will see below that this simple form in the mISS framework cannot explain the $1\sigma$ experimental  $(g-2)_{\mu}$ data  because there appears a strict correlation  with the decay rate Br$(\tau \to \mu \gamma)$ so that the experimental constraint results an upper bound on $\Delta a_{\mu,e}$.

\section{ \label{sec:Feynrules} Couplings and LFV analytical formulas}

\subsection{\label{subsec:LFVsource} LFV sources and $(g-2)_{e_a}$ anomalies}
All vertices  relating to couplings of singly charged Higgs bosons that give one-loop contributions to  $e_b \to e_a \gamma$ decay rates and $a_{e_a}$ anomalies are collected from the  Lagrangian \eqref{eq_ylepton1}. Namely, the relevant part in terms of physical lepton states is 
\begin{align}
\label{eq:LYlep}
	\mathcal{L}^Y_{\mathrm{lep}} =&- \overline{e_R}\frac{\sqrt{2}\hat{m}_{\ell}}{v_1} \left[ e_L \eta^0 + U^{e\dagger}_LU^{\nu}_0 n_L \left(-\eta^-\right) +   U^{e\dagger}_LV^{E}_L E_L \eta^{-A}  \right]
\crn & - \overline{E_R}\frac{\sqrt{2}\hat{m}_{E}}{v_3} \left[ V^{E \dagger}_L U^{e}_L e_L \chi^A + V^{E \dagger}_L U^{\nu}_0 n_L \left(-\chi^{-B}\right) +  E_L \chi^0  \right]
\crn & - \overline{n_R} \left( \frac{-\sqrt{2}}{v_2}\right) U^{\nu T}_1m_{D}\left[  U^{e}_L e_L \rho^+ + U^{\nu}_0 n_L \left(-\rho^0\right) +  V^{E}_L E_L \rho^{-B}  \right]
\crn &- \overline{n_L} U^{\nu \dagger}_2Y^h  U^{e}_R e_R h^+ +\mathrm{h.c.} +\dots, 
\end{align}  
where $\psi_{L(R)}=(\psi_1,\psi_2,\psi_3)^T_{L(R)}$ with $\psi=e,E$.  In the physical base  of Higgs and gauge bosons, Eq. \eqref{eq:LYlep} is   
\begin{align}
	\label{eq_g2Lagrangian}
	\mathcal{L}^Y_{\mathrm{lep}} =& -\frac{g}{\sqrt{2} m_W}\sum_{k=1}^2\sum_{a=1}^3 \sum_{i=1}^{3+2K} \overline{n_i} \left[ 	\lambda^{L,k}_{ia}  P_L +\lambda^{R,k}_{ia}P_R \right] e_a H^+_k   
	\crn &-\frac{g}{\sqrt{2} m_Y} \sum_{a,c=1}^3 V^{L*}_{ac}\overline{E_c} \left[ m_{E_c}t_{13}P_L + \frac{m_{e_a}}{t_{13}} P_R \right] e_a H^{A}\crn
	&+\sum_{a=1}^3\sum_{i=1}^{3+2K}\frac{g}{\sqrt{2}} \left( U^{\nu\dagger}_1U^e_L\right)_{ia} \overline{n_{i}}\gamma^{\mu}P_L e_aW^+_{\mu}   
	\crn &+\sum_{a,c=1}^3 \frac{g}{\sqrt{2}}  \left( V^{E\dagger}_{L}U^e_L\right)_{ca} \overline{E_c}\gamma^{\mu}P_L e_aY^{A}_{\mu} +\mathrm{h.c.},
\end{align}
where $t_{13}$ is defined in Eq. \eqref{eq_sij} and $\lambda^{L(R),k}_{i,a}$  calculated from  $U^{\nu}$ in Eq. \eqref{eq:Usimeq} are  
\begin{align}
	\label{eq_lakLR}
	\lambda^{L,1}_{ia}& = -t^{-1}_\beta c_{\alpha} \left( U_1^{\nu T} m_D U^e_L\right)_{ia}
	\crn %
	&\simeq \frac{t^{-1}_{\beta}c_{\alpha}}{ \sqrt{2}}\times \left[\begin{array}{cc}
		0,	& \quad i\leq 3 \\
i x_0^{\frac{1}{2}} \left( U^{*}_{\mathrm{PMNS}}  \hat{x}^{\frac{1}{2}}_{\nu}\xi^T\hat{\mu}_X^{-\frac{1}{2}}U^*_N \hat{M}_R\right)_{a(i-3)},	& \quad 0<i-3\leq K 
\\
- x_0^{\frac{1}{2}} \left( U^{*}_{\mathrm{PMNS}}  \hat{x}^{\frac{1}{2}}_{\nu}\xi^T\hat{\mu}_X^{-\frac{1}{2}}U^*_N \hat{M}_R\right)_{a(i-3-K)},	& \quad K<i-3\leq 2K  
	\end{array}\right.
	, 
	\crn	\lambda^{L,2}_{ia} &\simeq	\lambda^{L,1}_{ia}t_{\alpha},
	\crn	\lambda^{R,1}_{ia}& = -t_{\beta} m_a c_{\alpha} \left(U^{\nu \dagger}_0 U^{e}_{L}\right)_{ia} -\frac{vs_{\alpha}}{\sqrt{2}}  \left(U^{\nu \dagger}_2 Y^hU^e_R\right)_{ia}
\crn & \simeq  \left[\begin{array}{cc}
 	-m_{a}t_{\beta} c_{\alpha} \left(U^{\dagger}_{\mathrm{PMNS}} \right)_{ia} +  \frac{vs_{\alpha}}{\sqrt{2}}  \left(x_0^{\frac{1}{2}}\hat{x}_{\nu}^{\frac{1}{2}} \xi^{\dagger}\hat{\mu}_{X}^{-\frac{1}{2}} Y^hU^e_R\right)_{ia}  ,	& \quad i\leq 3 \\
  \frac{i \sqrt{x_0}}{\sqrt{2}}  m_{a}t_{\beta} c_{\alpha} \left( U^{\dagger}_{N} \hat{\mu}_X^{-\frac{1}{2}} \xi \hat{x}^{\frac{1}{2}}_{\nu} U^{\dagger}_{\mathrm{PMNS}}  \right)_{(i-3)a} +	\frac{ivs_{\alpha}}{2}\left(  U_N^{\dagger} Y^{h}U^e_R\right)_{(i-3) a}	& \quad 0<i-3\leq K 
 	\\
  -\frac{ \sqrt{x_0}}{\sqrt{2}}  m_{a}t_{\beta} c_{\alpha} \left( U^{\dagger}_{N} \hat{\mu}_X^{-\frac{1}{2}} \xi \hat{x}^{\frac{1}{2}}_{\nu} U^{\dagger}_{\mathrm{PMNS}}  \right)_{(i-3-K)a}- 	\frac{vs_{\alpha}}{2}\left(  U_N^{\dagger} Y^{h}U^e_R\right)_{(i-3-K) a}	& \quad K<i-3\leq 2K  
 \end{array}\right.,
	\crn	\lambda^{R,2}_{ia}&=	\lambda^{R,1}_{ia} \left[ c_{\alpha} \to s_{\alpha}, s_{\alpha}\to -c_{\alpha}\right], 
\end{align}  
For convenience to estimate the dominant one-loop contributions to $\Delta a_{e_a}$,   parts proportional to $x_0$ are ignored; see a detailed calculation  with $K=3$ in Refs.  \cite{Hue:2021zyw, Hue:2021xzl}. 

The form factors $c^X_{(ab)R}$ relating to new one-loop contributions from  exchanging the $X$ boson to the $\Delta a_{e_a}$ and cLFV decays were introduced in Ref.~\cite{Crivellin:2018qmi}.  Formulas of   $c^X_{(ab)R}$  with  $X=H^A,W^\pm, Y^{\pm A}$ were introduced precisely in Ref.  \cite{Hue:2021zyw} for the general seesaw scheme, consistent with  those discussed in Ref. \cite{Hue:2017lak}. It was shown previously that $c^X_{(22)R}\ll \Delta a^{\mathrm{NP}}_{\mu}$ with  $X=H^A,Y$ \cite{Hue:2021zyw};   we therefore fix  $c^X_{(ab)R}=0$ by a simple assumption that  $m_{E_a}\equiv m_E\; \forall a=1,2,3$, and $V^E_{L,R}=I_3$.   The contributions from singly charged Higgs and $W$ bosons are 
\begin{align}
	\label{eq:cabR}
c_{(ab)R} \left(H^\pm_k\right)=& \frac{g^2e\;}{32 \pi^2 m^2_Wm^2_{H^\pm_k} }  
	\crn&\times\sum_{i=1}^{3+2K}  \left[ \lambda^{L,k*}_{ia } \lambda^{R,k}_{ib }m_{n_i} f_{\Phi}(t_{i,k}) + \left( m_{b} \lambda^{L,k*}_{ia } \lambda^{L,k}_{ib } + m_{a} \lambda^{R,k*}_{ia } \lambda^{R,k}_{ib }\right)  \tilde{f}_{\Phi}(t_{i,k}) \right],
\crn c_{(ab)R} \left(H^\pm\right)=& \sum_{k=1}^2 c_{(ab)R} \left(H^\pm_k\right),
\crn 
c_{(ab)R}(W) \simeq& \frac{e G_Fm_{b}}{4\sqrt{2} \pi^2}   \left\{  -\frac{5}{12}\left[ \delta_{ab} - x_0 \left(U_{\mathrm{PMNS}}\hat{x}_{\nu}^{\frac{1}{2}}\xi^{\dagger}\hat{\mu}_X^{-1} \xi \hat{x}_{\nu}^{\frac{1}{2}} U_{\mathrm{PMNS}}^{\dagger} \right)_{ab} \right]
\right. \crn&\left. \hspace{1.4cm}+   x_0  \sum_{I=1}^K \left(U_{\mathrm{PMNS}} \hat{x}_{\nu}^{\frac{1}{2}}  \xi^\dagger  \hat{\mu}_X^{-\frac{1}{2}}U_N \right)_{aI} \left(U_N^{\dagger} \hat{\mu}_X^{-\frac{1}{2}} \xi \hat{x}_{\nu}^{\frac{1}{2}} U_{\mathrm{PMNS}}^\dagger \right)_{Ib} \tilde{f}_V \left(x_{I,W}\right)  \right\},
\end{align}
where $t_{i,k}\equiv m^2_{n_i} /m^2_{H^\pm_k}$, $x_{I,W} \equiv M^2_{I} /m^2_{W}$;   $U_{\mathrm{PMNS}}$ was defined in Eq. \eqref{eq:Upmns},  and the functions $f_{\Phi}(x)$, $\tilde{f}_{\Phi}(x)$, and $\tilde{f}_{V}(x)$ are given in Appendix \ref{app:ebagaloop}. We comment here that $c_{(ab)R}(W) $ consist of two parts, in which the first part proportional to $\delta_{ab}$ is exactly the SM results, and this first part does not contribute to LFV decay. In contrast,  the second part has a factor $x_0\ll1$; therefore this part gives small contribution to $\Delta a_{e_a}$, but still significantly affects  to Br$(\mu \to e\gamma)$ under the current experimental constraint. Therefore, we still add this part  in the numerical investigation.  In this work, we use    the limit $m_{n_i}=0$ with $i=1,2,3$; $m_{n_{3+I}}= m_{n_{3+K+I}}=M_I\; \forall \; I =\overline{1,K}$. The dominant contribution to $\Delta a_{e_a}$ is 
\begin{align}
	c_{(ab)R,0} \left(H^\pm_k\right) = \frac{g^2e vt^{-1}_{\beta}c_{\alpha}s_{\alpha}x_0^{\frac{1}{2}}}{32 \sqrt{2}\pi^2 m^2_W }   \sum_{I=1}^{K} \left[  \left(U_{\mathrm{PMNS}} \hat{x}^{1/2}_{\nu} \xi^{\dagger}\hat{\mu}_X^{-1/2}U_N \hat{M}_R  \right)_{aI} \left(U^{\dagger}_N Y^{h} U^e_R\right)_{Ib}  \frac{x_{I,k}f_{\Phi}(x_{I,k})}{ M_I}
	%
	  \right],\nn 
\end{align}
where  $U^e_R$ was defined  in Eq. \eqref{eq:UeLR};  $x_{I,k}\equiv M_I^2/m^2_{H^\pm_k}$ for  $k=1,2$; and $I=\overline{1,K}$. Consequently, the dominant contribution from the sum of the two singly charged Higgs bosons to $\Delta a_{e_a}$ [Br$(e_b\to e_a \gamma)$] when $a=b$ ($a\neq b$) is 
\begin{align}
c_{(ab)R,0} \left(H^\pm\right)& =  \frac{g^2e}{32 \pi^2 m^2_W }  \left[  \frac{vt^{-1}_{\beta} x_0^{\frac{1}{2}} c_{\alpha}s_{\alpha}}{\sqrt{2} }\left(U_{\mathrm{PMNS}} \hat{x}^{1/2}_{\nu} \xi^{\dagger}\hat{\mu}_X^{-1/2} U_N\hat{F}U^\dagger_N Y^{h} U^e_R\right)_{ab} \right] , 	\label{eq:cabRH1}
	\\  \hat{F}&\equiv \mathrm{diag}\left( \hat{F}_1, \; \hat{F}_2,\dots, \hat{F}_K\right), \label{eq:hatF}
	\crn  \hat{F}_I &= \sum_{k=1}^2(-1)^{k-1}x_{I,k}f_{\Phi}(x_{I,k})= x_{I,1}f_{\Phi}(x_{I,1}) -x_{I,2}f_{\Phi}(x_{I,2}). 
\end{align}
We emphasize that Eq. \eqref{eq:cabRH1} is the most general form, including the mixing matrices of charged and neutral leptons $e_a$ and $n_i$, and consistent with the result given in Ref. \cite{Hue:2021zyw} with $K=3$, $U^e_L=U^e_R=U_N=\xi= I_3$, and $M_I=M_0\; \forall I=\overline{1,K}$.  Equation \eqref{eq:cabRH1} is  very important for estimating analytic relations to cancel the large contributions to Br$(e_b\to e_a\gamma)$ consistent with experiments, while  keeping the large $\Delta a_{e,\mu}$. The numerical scanning time to determine the allowed points reduce dramatically. 

The one-loop contribution of  $X=H^\pm_{1,2},H^A,W,Y$ to $\Delta a_{e_a}$ is $	a_{e_a}(X)=- \frac{4m_{a}}{e}\mathrm{Re} \left[ 	c_{(aa)R}(X)\right]$,  leading to  a  deviation from the SM   as
\begin{align}
	\Delta a_{e_a}= &\sum_{X}  a_{e_a}(X) -a^{\mathrm{SM}}_{e_a}(W) \simeq - \frac{4m_{a}}{e}\mathrm{Re} \left[ 	c_{(aa)R}(H^\pm)\right],
\end{align}
where $a^{\mathrm{SM}}_{e_a}(W) $ is the one-loop contributions of $W$ boson predicted by the SM.  The heavy neutral Higgs and gauge  bosons give  suppressed one-loop  contributions to $\Delta a_{e_a}$ \cite{Hue:2021zyw}, so we omit them here.

\subsection{\label{sec:Feynman} One-loop contributions to LFV  decays}
From the information shown in Eqs. (29) and (30), we obtain all vertices needed for the  calculations all one-loop contributions to $e_b\to e_a \gamma$  in the unitary gauge, consistent with Refs. \cite{Hue:2017lak, Hue:2021zyw}. They are listed in Table \ref{t:hl331iSS}, 
\begin{table}[h]
	\begin{tabular}{|c|c|c|c|}
		\hline
		Vertex & Coupling & Vertex & Coupling \\
		\hline
		$\overline{n}_{i}e_{a}H^+_k$ & $ -\fr{ig}{\sqrt{2}m_W}\left( \lambda^{L,k}_{ia} P_L +\lambda^{R,k}_{ia} P_R\right) $ &
		$\overline{e_a}n_iH^-_k$ & $ -\fr{ig}{\sqrt{2}m_W}\left( \lambda^{R,k*}_{ia} P_L +\lambda^{L,k*}_{ia} P_R\right) $ \\
		\hline
		$\overline{E}_ce_a H^{+A}$&$\fr{-ig(V^{E\dagger}_{L}U^e_L)_{ca}}{\sqrt{2}m_Y}  \left( m_{E_c} t_{13} P_L +\frac{m_{a}}{t_{13}} P_R\right)$
		&$\overline{e}_aE_cH^{-A} $&$\fr{-ig (V^{E\dagger}_{L}U^e_L)^*_{ca}}{\sqrt{2}m_Y} \left( \frac{m_{a}}{t_{13}}  P_L+ m_{E_c} t_{13}  P_R\right)$\\
		\hline
		$\overline{n_{i}}e_{a}W^{+\mu}$ &$\frac{ig}{\sqrt{2}} (U^{\nu\dagger}_0U^e_L)_{ia}\gamma_\mu P_L $ &$\overline{e_a}n_i W^{-\mu}$
		& $\frac{ig}{\sqrt{2}}(U^{\nu\dagger}_0U^e_L)^*_{ia}\gamma_\mu P_L $\\
		\hline
		$\overline{E_c}e_a Y^{+A\mu}$ &$\frac{ig(V^{E\dagger}_{L}U^e_L)_{ca}}{\sqrt{2}}\gamma_\mu P_L$ &$\overline{e_a}E_cY^{-A\mu} $
		& $\frac{ig (V^{E\dagger}_{L}U^e_L)_{ca}^* }{\sqrt{2}}\gamma_\mu P_L$\\ 
		\hline
		$A^{\la}W^{+\mu}W^{-\nu}$&$-ie\Gamma_{\la \mu \nu}(p_0,p_{+},p_{-}) $&$A^{\la}Y^{+A\mu}Y^{-A\nu}$&$ -ieA \Gamma_{\la\mu\nu}(p_0,p_{+},p_{-})$\\
		\hline
		$A^{\mu}H^+_kH^-_k$&$ie(p_{+}-p_{-})_\mu$&$A^{\mu}H^{+A}H^{-A}$&$ieA(p_{+} -p_{-})_\mu$\\
		\hline
		$A^{\mu}\overline{e_a}e_a$&$-ie\gamma_\mu $&$A^{\mu}\overline{E_a} E_a$&$ieB\gamma_\mu  $\\
		\hline
	\end{tabular}
	\caption{Feynman rules for cLFV decays $e_b\to e_a \gamma$ in the 331$\beta$ISS$(K,K)$ model, $k=1,2$. 
	\label{t:hl331iSS}}
\end{table}
 following  conventions for general Lagrangian parts to derive these Feynman rules  given in Appendix \ref{app_Zeba}.  In the 331$\beta$ model, the two first components of the Higgs triplets can play roles of those appearing in the 2HDM; see a detailed discussion in Ref. \cite{Cherchiglia:2022zfy}, for example. As a result,  many available properties of the Higgs couplings and masses from the 2HDM model are valid in the 331$\beta$ model framework. On the other hand,  there are  $SU(3)_L$ particles such as   $H^A$, $Y^A$, and heavy charged leptons $E_a$ having masses in the $SU(3)_L$ scale $v_3$.  Studies of  searching for them  in colliders were discussed in Refs. \cite{Calabrese:2023ryr, Calabrese:2021lcz, CiezaMontalvo:2006zt}, which  suggest promising signals  in the future colliders if they exist. Note that in the 331$\beta$  the $SU(3)_L$ particles may carry exotic electric charges, implying the existence of long-lived exotic charged particles being searched for at the LHC \cite{CMS:2013czn}. Discussions for some particular channels relating to the $SU(3)_L$ lepton searches in 3-3-1 models are given in Refs. \cite{Suarez:2023ozu, Hue:2015mna, Pleitez:2021abk, Barreto:2023fye}. Constraints of lower bounds for these lepton masses are at least 600 GeV for charged leptons $E_a$ \cite{CMS:2024qys, ATLAS:2023zxo}.  The lower bound of $v_3$ was derived the most strictly from studying the production and decay of the heavy neutral gauge boson $Z'$ at the LHC and future colliders. Consequently, the mass of  the gauge boson $Y$ must be of the order of TeV;  therefore this boson gives suppressed one-loop contributions to the LFV decay amplitudes under consideration.

Here we use the general results given in Refs. \cite{Lavoura:2003xp, Crivellin:2018qmi, Hue:2017lak} in the limit of  heavy charged Higgs and gauge bosons, while tiny neutrino masses are still allowed.   The cLFV decays $e_b\to e_a\gamma$ will be used to constrain the parameter space of the model. The contributions to the decay rates are derived from the couplings and Feynman rules given in Table \ref{t:hl331iSS}, namely, 
\begin{align}
	\label{eq_Brebaga}
	\mathrm{Br}(e_b\to e_a \gamma)= \frac{48\pi^2}{m^2_{b}G_F^2 }\left( \left|c_{(ab) R}\right|^2 +\left|c_{(ba) R}\right|^2\right) \mathrm{Br}(e_b\to e_a\overline{\nu_a}\nu_b),
\end{align}
where $G_F=g^2/(4\sqrt{2}m_W^2)$,   and 
\begin{align}
	c_{(ab)R} &= c_{(ab)R}(W) + 	c_{(ab)R}(Y) +c_{(ab)R}(H^\pm) +c_{(ab)R}({H^A}). 
 	\label{eq:cabR}
\end{align} 
Apart from  Table \ref{t:hl331iSS}, the couplings for one-loop contributions to the LFV$h$ decays $h\to e_ae_b$ are listed as the following. 

From the above discussion on the Higgs potential, we can derive all  Higgs self-couplings of the SM-like Higgs boson  using the interacting Lagrangian $$\mathcal{L}_{hHH}=-V_h =\sum_{s_1s_2} \left( -\lambda_{hs_1s_2} hs^Q_1s_2^{-Q} +\mathrm{h.c.}\right) +\dots.$$ The Feynman rule for a vertex $ hs^Q_1s_2^{-Q}$ is $(-i\lambda_{hs_1s_2})$, where  $s_1,s_2=H^{\pm}_{1,2},H^{\pm A}$ and 
\begin{align}
	\label{eq:hss}
	\lambda_{h H^+_1 H^-_1}=& 2 c_{\alpha }^2 c_{\beta } c_{\zeta } \lambda _1 s_{\beta }^2 v -2 c_{\alpha }^2 c_{\beta }^2 \lambda _2 s_{\beta } s_{\zeta } v + c_{\alpha }^2 \lambda _{12} v \left(c_{\beta }^3 c_{\zeta }-s_{\beta }^3 s_{\zeta }\right) +c_{\alpha }^2 c_{\delta } \tilde{\lambda }_{12} v
	\crn & +s_{\alpha }^2 v (c_{\beta } c_{\zeta } \lambda _2^h-s_{\beta } s_{\zeta } \lambda _1^h) -\sqrt{2} c_{\alpha } c_{\delta } f_h s_{\alpha },
	\crn 
	\lambda_{h H^+_2 H^-_2}=& 2 c_{\beta } s_{\alpha }^2 s_{\beta } v (c_{\zeta } \lambda _1 s_{\beta }-c_{\beta } \lambda _2 s_{\zeta }) +\lambda _{12} s_{\alpha }^2 v \left(c_{\beta }^3 c_{\zeta }-s_{\beta }^3 s_{\zeta }\right) +c_{\delta } \tilde{\lambda }_{12} s_{\alpha }^2 v
	\crn&+ c_{\alpha }^2 v (c_{\beta } c_{\zeta } \lambda _2^h-s_{\beta } s_{\zeta } \lambda _1^h) +\sqrt{2} c_{\alpha } c_{\delta } f_h s_{\alpha },
	\crn 
	\lambda_{h H^\pm_1 H^\mp_2}=& 2 c_{\alpha } c_{\beta } s_{\alpha } s_{\beta } v (c_{\zeta } \lambda _1 s_{\beta }-c_{\beta } \lambda _2 s_{\zeta }) +c_{\alpha } \lambda _{12} s_{\alpha } v \left(c_{\beta }^3 c_{\zeta }-s_{\beta }^3 s_{\zeta }\right) +c_{\alpha } c_{\delta } \tilde{\lambda }_{12} s_{\alpha } v 
	\crn& +c_{\alpha } s_{\alpha } v (s_{\beta } s_{\zeta } \lambda _1^h-c_{\beta } c_{\zeta } \lambda _2^h) +\frac{c_{\delta } f_h \left(c_{\alpha }^2-s_{\alpha }^2\right)}{\sqrt{2}},
\crn 	\lambda_{h H^A H^{-A}}=&c_{13}^2 v (2 c_{\beta } c_{\zeta } \lambda _1-\lambda _{12} s_{\beta } s_{\zeta }) +s_{13}^2 v (c_{\beta } c_{\zeta } \lambda _{13}-\lambda _{23} s_{\beta } s_{\zeta }) +c_{\zeta } \tilde{\lambda }_{13} s_{13} (c_{13} v_3+c_{\beta } s_{13} v)
\crn &+2 f c_{13} s_{13} s_{\zeta }.
\end{align}

The couplings of Higgs and gauge bosons are contained in the covariant kinetic terms of the Higgs bosons. From general notations discussed  in Appendix \ref{app_Zeba}, the  Feynman  rules for particular couplings  relating to SM-like Higgs boson are derived using the vertex factors  shown in Table~\ref{table_HGcoupling},
\begin{table}[ht]
\centering 
\begin{tabular}{|c|c|c|c|}
	\hline 
	Vertex	& Coupling: & Vertex&Coupling\\ 
	\hline 
	$g_{h W^+W^-}$	&$g \,m_W \,c_{\delta}$& & \\
	\hline 
	  $g_{h H^-_2W^+}$ & $ -\frac{gc_{\alpha } s_{\delta }}{2} $ &  $g_{h H^-_1W^+}$ & $ -\frac{gs_{\alpha } s_{\delta }}{2} $ 	\\
	\hline 
 $g_{hY^{+A}Y^{-A}}$ &  $ g m_Wc_{\beta } c_{\zeta } $ &	$g_{h H^{-A}Y^{A}}$&  $-\frac{g\,c_{13}c_{\zeta}}{2} $ \\
	\hline
\end{tabular}
\caption{Vertex factors  for SM-like Higgs  couplings to charged Higgs and  gauge bosons in the  331$\beta$ISS$(K,K)$ model.} \label{table_HGcoupling}
\end{table}
which is consistent with Ref. \cite{Hung:2019jue} in the limit $\alpha=0$. The $H^\pm_k h W^\mp$ couplings with k=1,2 always consist of the suppressed factor $s_{\delta}\ll1$, which has the same property known in 2HDM. In particular, $s_{\delta}=0$ is the aligned limit to guarantee that the SM-like Higgs couplings are consistent with the SM prediction. The small $s_{\delta}$ is also consistent with the nonobservations of $H^\pm_{1}$ through the decay channel search $H^\pm\to hW^\pm$ at the LHC. Therefore, we can ignore the relevant one-loop contributions to the LFV$h$ decay amplitudes $h\to e_ae_b$ in numerical investigations.

Similarly, for the one-loop contributions to the decay $Z\to e_b^+e_a^- $,  the needed couplings are given in Ref. \cite{Hung:2019jue}; see Table \ref{table_Z1A}. 
\begin{table}[ht]
	\centering 
	\begin{tabular}{|c|c|}
	\hline
	Vertex& Coupling\\
	\hline
	$g_{ZH^+_1H^-_1}$	& $ \frac{1}{2c_Ws_W}\left[c_{\theta } \left(c_{\alpha }^2-2 s_W^2\right) +\frac{c_W s_{\theta } \left(c_{\alpha }^2 \left(2 \sqrt{3} s_{\beta }^2+3 \beta  t_W^2-\sqrt{3}\right)+6 \beta  s_{\alpha }^2 t_W^2\right)}{3 \sqrt{1-\beta ^2 t_W^2}} \right] $\\
	\hline 	
	$g_{ZH^+_2H^-_2}$	& $ \frac{1}{2s_Wc_W}\left[  c_{\theta } \left(s_{\alpha }^2-2 s_W^2\right) +\frac{c_W s_{\theta } \left(6 \beta  c_{\alpha }^2 t_W^2+s_{\alpha }^2 \left(2 \sqrt{3} s_{\beta }^2+3 \beta  t_W^2-\sqrt{3}\right)\right)}{3 \sqrt{1-\beta ^2 t_W^2}}\right] $ \\
	\hline 	
	$g_{ZH^+_{1(2)}H^-_{2(1)}}$	&  $\frac{s_{\alpha } c_{\alpha }}{2s_Wc_W}  \left[c_{\theta }  -\frac{ c_Ws_{\theta } \left(-2 \sqrt{3} s_{\beta }^2+3 \beta  t_W^2+\sqrt{3}\right)}{3 \sqrt{1-\beta ^2 t_W^2}}  \right] $\\
	\hline 	
	$g_{ZH^{A}H^{-A}}$	& $\frac{1}{2s_Wc_W}\left( c_{\theta} \left[ s^2_{13} -(1 +\sqrt{3}\beta)s_W^2 \right] +\frac{s_{\theta}\left[\sqrt{3}c^2_W(s^2_{13}-2) +3\beta (\sqrt{3}\beta +c^2_{13})  s^2_W\right]}{3c_W\sqrt{1-\beta^2t^2_W}} \right) $\\
	\hline 	
	$g_{ZW^{+}H^-_1}$& $-\frac{2  c_{\alpha } c_{\beta } m_W s_{\beta } s_{\theta }}{s_W \sqrt{3-3 \beta ^2 t_W^2}}$\\
	\hline 
	$g_{ZW^{+}H^-_2}$& $-\frac{2  c_{\alpha } c_{\beta } m_W s_{\beta } s_{\theta }}{2 s_W \sqrt{3-3 \beta ^2 t_W^2}}$\\
	\hline 
	$g_{ZY^{A}H^{-A}}$ & $ \frac{gc_{13}}{4s_W} \Big\{ c_{\theta}c_W \left[ s_{12}\left( 1 +(2 +\sqrt{3}\beta)t^2_W \right)v +t_{13}(1 -\sqrt{3}\beta t^2_W) v_3 \right] $\\
	& $\left. +\frac{s_{\theta}}{3\sqrt{1-\beta^2t^2_W}} \left[s_{12} \left(\sqrt{3} -3\beta(2 +\sqrt{3}\beta)t^2_W\right) v +\sqrt{3}t_{13} \left(1 +3 \beta^2t_W^2 \right)v_3 \right]\right\}$\\
	\hline 
\end{tabular}
	\caption{Feynman rules of couplings with $Z$ to charged Higgs and gauge bosons. }\label{table_Z1A}
\end{table}
Because $g_{ZW^+H^-_1}\varpropto s_{\theta}=\mathcal{O}(v^2/v_3^2)\leq \mathcal{O}(10^{-4})$, one-loop contributions to LFV$Z$ amplitudes relating to $ZW^\pm H^\mp_k$ are suppressed. 

The triple couplings of three gauge bosons arise from the covariant kinetic Lagrangian of the non Abelian gauge bosons leading to  the involved  couplings of $Z$  given in Table~\ref{table_3gaugcoupling}.
\begin{table}[ht]
	\centering \begin{tabular}{|c|c|}
		\hline
		Vertex & Coupling\\
		\hline 
		$g_{ZW^{+}W^{-}}$& $t_W^{-1}c_{\theta}
		$\\
		\hline 
		$g_{ZY^{A}Y^{-A}}$& $ \frac{  1}{2s_W} \left[  c_{\theta} c_W\left(-1	 +\sqrt{3} \beta  t^2_W\right) +s_{\theta} \sqrt{3 -3 \beta^2  t_W^2}\right] 
		$\\
		\hline 
	 	\end{tabular}
	\caption{Feynman rules for  triple gauge couplings  relating with LFV decays.
	} \label{table_3gaugcoupling}
\end{table}

The remaining couplings of the SM-like Higgs boson and $Z$ to two Dirac leptons are listed in Table \ref{tab_Bffp}.
\begin{table}[ht]
	\centering 
	\begin{tabular}{c c }
		\hline
		Vertex& Coupling\\
		\hline
		$h\overline{e_a}e_a$	&  $-\frac{im_{a}}{v}\times \frac{c_{\zeta}}{c_{\beta}}$\\
			\hline 
		$Z\overline{e_a}e_a$	&  $\frac{iec_{\theta}}{2c_Ws_W} \left[ \left(-1 +2 s_W^2 +\frac{t_{\theta}}{c_W \sqrt{1-\beta^2 t^2_W}}  \left( \frac{c^2_W}{\sqrt{3}} -\beta s^2_W\right)\right) P_L + 2\left( s^2_W - \frac{\beta s^2_Wt_{\theta} }{c_W \sqrt{1 -\beta^2 t^2_W}}\right) P_R\right]$   \\
		\hline 
		$Z\overline{E_a}E_a$	&  $\frac{igc_{\theta}}{2c_W} \left[ \left(  (1 -\sqrt{3}\beta)  s_W^2 + \frac{ 3 \beta (-1 +\sqrt{3} \beta) s^2_W t_{\theta}+ (-2\sqrt{3} c_W^2 )t_{\theta}}{3c_W \sqrt{1 -\beta^2 t^2_W}} \right)  + \left( \frac{  (-2\sqrt{3} c_W^2 )t_{\theta}}{3c_W \sqrt{1 -\beta^2 t^2_W}} \right)P_L  \right] $ \\
		\hline 
	\end{tabular}
	\caption{Feynman rules of couplings with $Z/h$ to charged  leptons. }\label{tab_Bffp}
\end{table}
It is noted that the couplings of the $h$ and $Z$ to two  Majorana fermions must be symmetric   for the right expressions for Feynman rules in notations of the four-components Dirac spinor \cite{Dreiner:2008tw}. This is the case of  neutrinos with masses generated following the  seesaw mechanism.  The Feynman rules now  for couplings of a neutral  bosons $Z$ (or $h$) to two Majorana fermions  are presented in Fig. \ref{tab_BffM}, 
\begin{table}[ht]
	\centering 
	\begin{tabular}{c c }
		\hline
		Vertex& Coupling\\
		\hline
		$h\overline{n_i}n_j$	&  $ -\frac{ig}{2m_W} \left(\lambda^h_{ij} P_L + \lambda^{h*}_{ij} P_R \right)$  \\
		\hline 
		$Z_{\mu}\overline{n_i}n_j$	&   $ ie \gamma^{\mu}\left(  G_{ij} P_L -G_{ji} P_R  \right)$\\
	\hline 
	\end{tabular}
	\caption{Feynman rules of couplings with $Z/h$ to two Majorana leptons $n_i$ and $n_j$. }\label{tab_BffM}
\end{table}
corresponding to the following Lagrangian 
\begin{align}
\label{eq_LintM}
\mathcal{L}_{\mathrm{int}}= \sum_{i,j=1}^{3+2K} \left[ -\frac{g}{4m_W} \overline{n_i} \left( \lambda^h_{ij} P_L  + \lambda^{h*}_{ij} P_R\right) n_j h  + \frac{e}{2}\overline{n_i}\gamma^{\mu}\left( G_{ij} P_L  -G_{ji} P_R\right) n_j Z_{\mu}\right],
\end{align}
where 
\begin{align}
	\label{eq:Gij}
\lambda^h_{ij}=& \lambda^h_{ji} =-\frac{s_{\zeta}}{s_{\beta}} \sum_{c=1}^3 \left( U^{\nu*}_{ci}m_{n_i}U^{\nu}_{cj} + U^{\nu*}_{cj}m_{n_j}U^{\nu}_{ci}\right) ,
\\		G_{ij}= & \frac{c_{\theta}}{2c_Ws_W} \left[ 1+\frac{t_{\theta}}{c_W \sqrt{1-\beta^2 t^2_W}}  \left( \frac{c^2_W}{\sqrt{3}} -\beta s^2_W\right)  \right] q_{ij},  \; q_{ij}\equiv \sum_{c=1}^3U^{\nu*}_{ci}U^{\nu}_{cj} . 
\end{align}
The relation $\frac{1}{2}\overline{n_i}\gamma^{\mu}\left( G_{ij} P_L  -G_{ji} P_R\right) n_j =\overline{n_i}\gamma^{\mu}G_{ij} P_Ln_j$ was used \cite{Dreiner:2008tw}. After deriving the Feynman rules for vertex couplings, the amplitudes will be written in the same  technique for both kinds of  Dirac and Majorana  fermions appearing in the vertices $Z\overline{n_i} n_j$ and $Z\overline{E_a}E_a$, respectively \cite{Dreiner:2008tw}.

\begin{figure}[ht]
	\centering 
	\includegraphics[width=15cm]{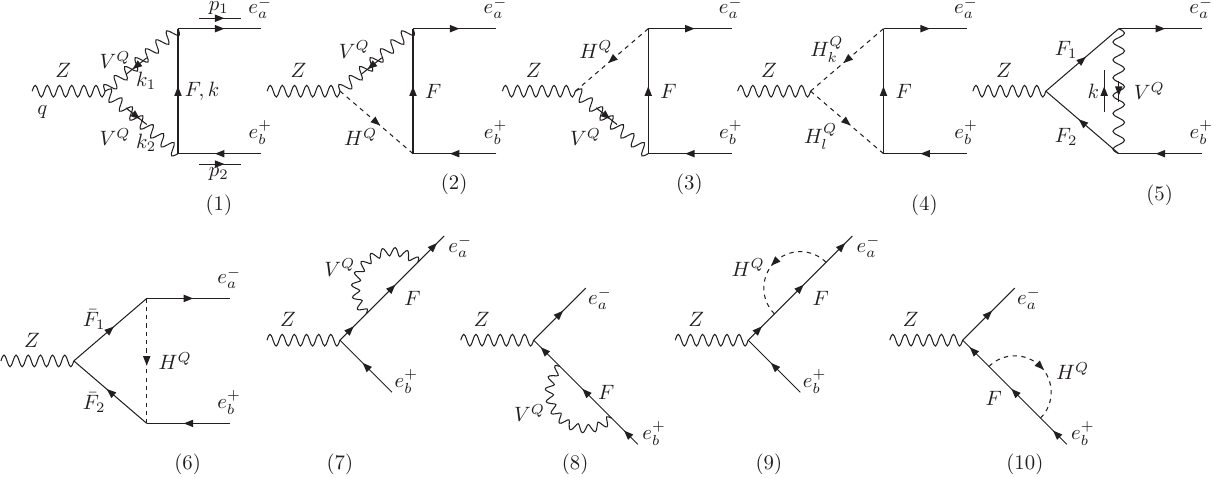}
	\caption{ Feynman diagrams giving  one-loop contributions to the decays $Z\rightarrow e_b^+ e_a^- $ in the unitary gauge, where $F,F_1,F_2\equiv n_i, E_a$;   $H,H_{k,l}=H^{\pm},H^{\pm A}$; and  $V=W^{\pm}, Y^{\pm A}$ in the 331$\beta$ISS$(K,K)$ model.}\label{fig_LFVZ}
\end{figure}

The one-loop Feynman diagram for  decays $Z\to e_a^\pm  e_b^\mp$ are shown in Fig. \ref{fig_LFVZ}.
We use the formulas of amplitude for the decay $Z(q)\to e^+_b(p_2) e^-_a(p_1)$ in Ref. \cite{Jurciukonis:2021izn} 
\begin{align}
\label{eq_Mzeab}
\mathcal{M}_Z& =  \frac{e}{16 \pi^2} \times \overline{u_a}\left[ \slashed{\varepsilon} \left( \bar{a}_LP_L + \bar{a}_R P_R\right)  + (p_1.\varepsilon) \left( \bar{b}_L P_L + \bar{b}_R P_R\right)\right] v_b,
\end{align}
where  $\varepsilon^{\mu}$ is the polarized vector of the gauge boson $Z$  and  $\bar{b}_{L,R}$.  The LFV$Z$ partial decay width now is 
\begin{align}
\label{eq_brLFVZ}
\Gamma(Z\to e^+_b e^-_a )= \frac{\sqrt{\lambda}}{16 \pi m_Z^3}\left| \frac{e}{16\pi^2}\right|^2 \left( \frac{\lambda H_0}{12 m_Z^2}  +H_1 + \frac{ H_2}{3 m_Z^2} \right),
\end{align} 
where 
\begin{align}
\label{eq_not1}
\lambda :=& m_Z^4  + m_a^4 + m_b^4 -2 \left( m_Z^2  m_a^2 + m_Z^2  m_b^2 + m_a^2 m_b^2\right),
\crn H_0 =& \left( m_Z^2 -m_a^2 -m_b^2\right) \left( |\bar{b}_L|^2 +  |\bar{b}_R|^2\right) - 4m_a m_b \mathrm{Re}\left[\overline{b}_L \overline{b}_R^*\right] 
\crn & - 4 m_a \mathrm{Re}\left[ \overline{a}_R^* \overline{b}_L +\overline{a}_L^* \overline{b}_R \right] - 4 m_b \mathrm{Re}\left[ \overline{a}_L^* \overline{b}_L  +\overline{a}_R^* \overline{b}_R \right],
\crn H_1 =& 4 m_a m_b  \mathrm{Re}(\overline{a}_L \overline{a}_R^*),
\crn  H_2= &  \left[ 2 m_Z^4 -m_Z^2 \left( m_a^2 +m_b^2\right) -  \left( m_a^2 -m_b^2\right)^2 \right] \left( |\overline{a}_L|^2 + |\overline{a}_R|^2\right),
\end{align}
where the form factors $\overline{a}_{L,R}$ arise from loop corrections corresponding to Feynman diagrams shown in Fig. \ref{fig_LFVZ}.   
The detailed analytic formulas  to $\bar{a}_{L,R}$ and $\bar{b}_{L,R}$ are shown in Appendix \ref{app_Zeba} based on previous works  \cite{Jurciukonis:2021izn, Abada:2022asx, Hong:2023rhg}. On the other hand, a new class of one-loop contributions from $Z$-$S$-$V$ diagrams 2 and 3 was computed precisely. 

\begin{figure}[ht]
	\includegraphics[width=15cm]{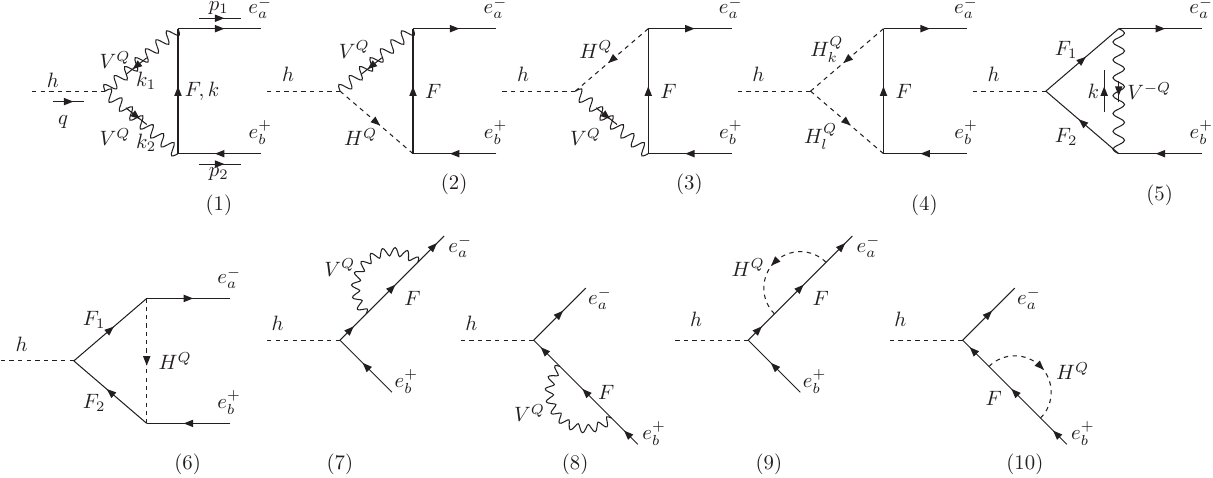}
	\caption{Feynman diagrams giving  one-loop contributions  to the decay $h\rightarrow e^-_ae^+_b$ in the unitary gauge, where $F,F_1,F_2\equiv n_i, E_a$;   $H,H_{k,l}=H^{\pm},H^{\pm A}$; and  $V=W^{\pm}, Y^{\pm A}$.}\label{fig_LFVh}
\end{figure}

In the unitary gauge, the Feynman diagrams for one-loop contributions to LFV$h$ decays are shown in Fig. \ref{fig_LFVh}   \cite{Arganda:2004bz, Arganda:2014dta, Nguyen:2018rlb, Hong:2022xjg, Hue:2015fbb}.  
The effective Lagrangian of the  decays  is
$$ \mathcal{L}^{\mathrm{LFV}h}= h \overline{e_a} \left[\Delta_{(ab)L} P_L  +\Delta_{(ab)R}  P_R \right]  e_b + \mathrm{H.c.},$$
where the scalar factors $\Delta_{(ab)L,R}$  arise from the loop contributions.  
The partial width  is
\begin{equation}
	\Gamma (h\rightarrow e_ae_b)\equiv\Gamma (h \rightarrow e_a^{-} e_b^{+})+\Gamma (h\rightarrow e_a^{+} e_b^{-})
	\simeq   \fr{ m_{h} }{8\pi }\left(\vert \Delta_{(ab)L}\vert^2+\vert \Delta_{(ab)R}\vert^2\right) 
	\label{LFVwidth}
\end{equation}
with the condition  $m^2_{h}\gg m^2_{a,b}$. Here, $m_{a,b}$ are the masses of the charged leptons. The on-shell conditions for external particles are $p^2_{1,2}=m_{a,b}^2$  and $ q^2 \equiv( p_1+p_2)^2=m^2_{h}$. The corresponding branching ratio is  Br$(h \rightarrow e_ae_b)= \Gamma (h\rightarrow e_ae_b)/\Gamma^{\mathrm{total}}_{h},$ where $\Gamma^{\mathrm{total}}_{h}\simeq 4.1\times 10^{-3}$ GeV \cite{Denner:2011mq}.  The $\Delta_{(ab)L,R}$ can be written as
\be \Delta_{(ab)L,R} = \sum_{i=1,5,7,8} \Delta^{(i)W}_{(ab)L,R}   + \sum_{i=4,6,9,10} \Delta^{(i)H}_{(ab)L,R},  \label{deLR}\ee
where the analytic forms of  $\Delta^{(i)W}_{(ab)L,R}$ and $\Delta^{(i)H}_{(ab)L,R}$ are  shown in  Appendix \ref{app:heba}. We recall here that one-loop contributions from $E_a,H^{\pm A}$,  and $Y$ are zeros with assumptions that $V^E_{L,R}=I_3$ and $m_{E_a}=m_E\;\forall a=1,2,3$.

\section{\label{sec_numerical} Numerical discussion}

In numerical discussion, we will use the best-fit values  of the neutrino oscillation data~\cite{Workman:2022ynf} corresponding to  the NO scheme  with $m_{n_1}<m_{n_2}<m_{n_3}$, namely,  
\begin{align}
	\label{eq_d2mijNO}
	&s^2_{12}=0.32,\;   s^2_{23}= 0.547,\; s^2_{13}= 0.0216 ,\;  \delta= 218 \;[\mathrm{Deg}] , 
	\crn &\Delta m^2_{21}=7.55\times 10^{-5} [\mathrm{eV}^2], \;
	\Delta m^2_{32}=2.424\times 10^{-3} [\mathrm{eV}^2].
\end{align}
In numerical calculation,   we  will use the  formulas 
\begin{align}
	\hat{m}_{\nu}&=  \left( \hat{m}^2_{\nu}\right)^{1/2}= \mathrm{diag} \left( m_{n_1}, \; \sqrt{ m_{n_1}^2 + \Delta m^2_{21}},\; \sqrt{m_{n_1}^2  +\Delta m^2_{21} +\Delta m^2_{32}} \right), 	\label{eq:NOmnu}
	\\  U_{\mathrm{PMNS}} &=\left(
	\begin{array}{ccc}
		c_{12} c_{13} & c_{13} s_{12} & s_{13} e^{-i \delta } \\
		-c_{23} s_{12}-c_{12} s_{13} s_{23} e^{i \delta } & c_{13} c_{23}-s_{12} s_{13} s_{23} e^{i \delta } & c_{13} s_{23} \\
		s_{12} s_{23}-c_{12} c_{23} s_{13} e^{i \delta } & -c_{23} s_{12} e^{i \delta } s_{13}-c_{13} s_{23} & c_{13} c_{23} \\
	\end{array}
	\right). 	\label{eq:NOupmns}
\end{align}

Similarly for the IO scheme with   $m_{n_3}<m_{n_1}<m_{n_2}$, we choose experimental data   as follows \cite{Workman:2022ynf}: 
\begin{align}
	\label{eq_d2mijNO}
	&s^2_{12}=0.318^{+0.016}_{-0.016},\;   s^2_{23}= 0.578_{-0.010}^{+0.017},\; s^2_{13}= 2.225^{+0.064}_{-0.070} \times 10^{-2} ,\;  \delta= 284^{+26}_{-28} \;[\mathrm{Deg}] , 
	\crn &\Delta m^2_{21}=7.5^{+0.22}_{-0.20} \times 10^{-5} [\mathrm{eV}^2], \;
	\Delta m^2_{32}=-2.52^{+0.03}_{-0.02}\times 10^{-3} [\mathrm{eV}^2]. 
\end{align} 
The active mixing matrix and neutrino masses are determined  as 
\begin{align}
	\label{eq_IO}
	\hat{m}_{\nu}&= \left( \hat{m}^2_{\nu}\right)^{1/2}= \mathrm{diag} \left( \sqrt{m_{n_3}^2 -\Delta m^2_{32} -\Delta m^2_{21}}, \;\sqrt{m_{n_3}^2 -\Delta m^2_{32}},\; m_{n_3}  \right). 
\end{align}
These neutrino masses satisfy automatically the constraint  from Plank 2018 \cite{Planck:2018vyg} that   $\sum_{i=1}^{3}m_{n_i}\leq 0.12\; \mathrm{eV}$. This condition is automatically satisfied in the mISS framework, $m_{n_1}=0$ ($m_{n_3}=0$) for the NO (IO) scheme. In general, the upper bound of the lightest active neutrino mass if $m_{n_1}(m_{n_3})<0.031$ eV in the NO (IO) scheme.  

The other well-known numerical parameters are given in Ref.~\cite{Workman:2022ynf}, namely, 
\begin{align}
	\label{eq_ex}
	g &=0.652,\; G_F=1.166\times 10^{-5} \;\mathrm{GeV}^{-2},\;\alpha_e=\frac{1}{137}= \frac{e^2}{4\pi} ,\; s^2_{W}=0.231, 
\crn  m_W& =80.377 \; \mathrm{GeV},\;  m_Z =91.1876 \; \mathrm{GeV},\; m_h=125.25\; \mathrm{GeV}, \Gamma_Z=2.4955 \; \mathrm{GeV}, 
\crn 	m_e&=5\times 10^{-4} \;\mathrm{GeV},\; m_{\mu}=0.105 \;\mathrm{GeV} ,\; m_{\tau}=1.776 \;\mathrm{GeV}. 
\end{align}
The general 331$\beta$ model predicts a  $SU(3)_L$ scale $v_3$ indicated to be as large as the order of $\mathcal{O}(10)$ TeV \cite{Cherchiglia:2022zfy, Okada:2016whh, deJesus:2023lvn, Long:2018fud, Alves:2022hcp, Suarez:2023ozu}. Therefore,  we will fix $v_3=20$ TeV, leading to the suppressed $Z-Z'$ mixing angle $\theta\simeq 0$.  For simplicity, we fix $\theta=0$; therefore all parts relating to $\beta$ appearing in the coupling factors vanish.   Recall that  one-loop contributions from $SU(3)_L$ charged Higgs $H^A$ and gauge bosons $Y$  with  $m_Y>1$ TeV and $m_Ec_{\beta}>5$ GeV result in the largest values of $ \Delta a_{\mu}(H^A) \leq \mathcal{O}(10^{-11})$ \cite{Hue:2021zyw}. These bosons also give zero one-loop contributions to LFV decay amplitudes in the case of $V^E_{L,R}=I_3$; hence they do not play significant roles in our work.

The nonunitary part $\eta\equiv	\frac{1}{2}\left| R_0R_0^{\dagger}\right| \varpropto x_0$ of the active neutrino mixing matrix is constrained from studies on  electroweak precision, cLFV decays \cite{Fernandez-Martinez:2016lgt, Agostinho:2017wfs, Blennow:2023mqx}. 
   We will choose the values that  $x_0=|\eta_{33}| \leq 10^{-4}$ in our numerical discussion, consistent with Refs. \cite{Biggio:2019eeo, Escribano:2021css, Pinheiro:2021mps}. 

It was shown that heavy charged Higgs bosons still allow  small $t_{\beta}$ \cite{Okada:2016whh, Sanchez-Vega:2018qje, Cherchiglia:2022zfy, Botella:2022rte}, which supports the $(g-2)_{e_a}$ data in the models under consideration. The heavy neutrino masses must satisfy the experimental searches \cite{Das:2012ze, Das:2017nvm, Das:2017pvt, Bhardwaj:2018lma}. Therefore, we scan the following regions of the parameter space:
\begin{align}
	\label{eq_scanX}
&t_{\beta}\in [0.5,20]; \; \alpha \in [0,2\pi];\; x_0\in [10^{-6},10^{-4}];\;M_I \in [1,20] \;(\mathrm{TeV}); \; m_{H^\pm_{1,2}} \in [1,5] \;(\mathrm{TeV}); 
\crn & \mathrm{max}\left[|Y^h_{Ia}|\right],\; \mathrm{max}\left[|Y^\nu_{Ia}|\right]<\sqrt{4\pi}\; \forall I=1,2,\dots, K; a=1,2,3. 
\end{align}
We also fix $\zeta=-\theta_{21}$ corresponding to $\delta=0$ to get the SM prediction for Higgs decays consistent with experiments. For investigating the LFV$h$ decay, we  apply allowed conditions  of Higgs self-couplings discussed in Refs. \cite{Costantini:2020xrn, Li:2021poy, Cherchiglia:2022zfy} summarized in Appendix \ref{app_Higgs}.

\begin{figure}[ht]
%
\begin{tabular}{ccc}
\includegraphics[width=5.5cm]{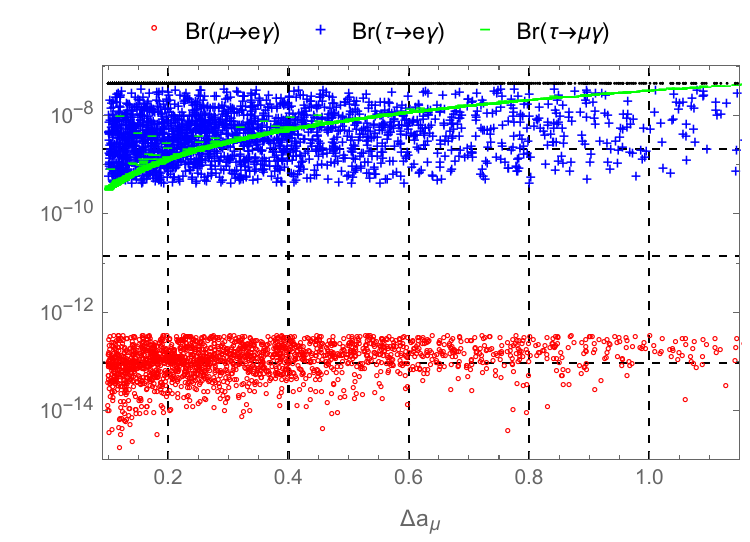}	&  
\includegraphics[width=5.5cm]{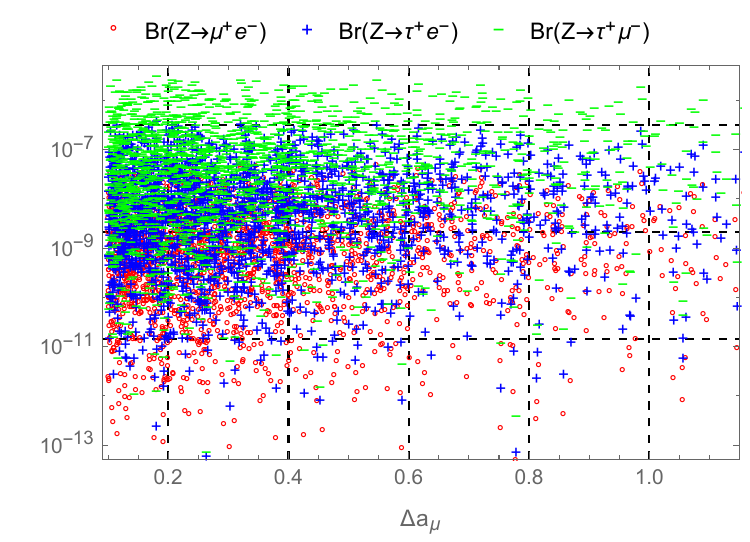}	& 
\includegraphics[width=5.5cm]{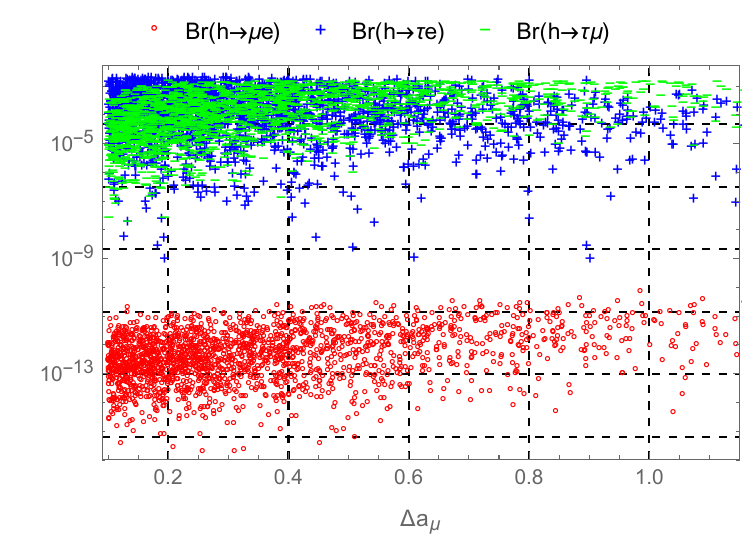}	 
\end{tabular}
\caption{ LFV decay rates as functions of $\Delta a_{\mu}$ predicted by the 331$\beta$mISS model (NO). The black line in the left panel shows the constant value of $4.2\times 10^{-8}$. The narrow green curve in the left panel shows the strict and nearly linear  dependence of Br$(\tau \to \mu \gamma)$ on $\Delta a_{\mu}$.  The remaining  LFV rates do not have this property.}\label{fig_amLFVm}
\end{figure}

First, we consider the 331mISS model with $K=2$.  Before showing the precisely numerical results, we estimate the effects of constraints  of  Br$(e_b\to e_a \gamma)$ on $\Delta a_{\mu,e}$ using Eq. \eqref{eq:cabRH1}, with well-known  experimental data of $\Delta a_{e,\mu}$. Fixing  $c_{(12)R,0} \left(H^\pm\right)=c_{(21)R,0} \left(H^\pm\right)=0$ to keep small Br$(\mu\to e\gamma)$, there are strict relations between $\Delta a_{e,\mu}$ and Br$(e_b \to e_a \gamma)$ independent with all entries of $Y^h$. Namely,  the NO scheme predicts that
\begin{align}
	\label{eq:fR12}
\left| \frac{c_{(22)R,0} \left(H^\pm\right)}{c_{(32)R,0} \left(H^\pm\right)} \right|	(\mathrm{NO}) & = \left|\frac{\left( U_{\mathrm{PMNS}}\right)_{13} \left( U_{\mathrm{PMNS}}\right)_{22} -\left( U_{\mathrm{PMNS}}\right)_{12} \left( U_{\mathrm{PMNS}}\right)_{23}}{\left( U_{\mathrm{PMNS}}\right)_{22} \left( U_{\mathrm{PMNS}}\right)_{33} -\left( U_{\mathrm{PMNS}}\right)_{23} \left( U_{\mathrm{PMNS}}\right)_{32}} \right| \simeq 1.8,
\crn 
\left| \frac{c_{(11)R,0} \left(H^\pm\right)}{c_{(31)R,0} \left(H^\pm\right)}\right|  (\mathrm{NO}) & = \left| \frac{\left( U_{\mathrm{PMNS}}\right)_{13} \left( U_{\mathrm{PMNS}}\right)_{22} -\left( U_{\mathrm{PMNS}}\right)_{12} \left( U_{\mathrm{PMNS}}\right)_{23}}{\left( U_{\mathrm{PMNS}}\right)_{13}  \left( U_{\mathrm{PMNS}}\right)_{32} -\left( U_{\mathrm{PMNS}}\right)_{12} \left( U_{\mathrm{PMNS}}\right)_{33}}\right| \simeq 0.62, 
\end{align}
where the numerical values corresponding to the best-fit point of  $U_{\mathrm{PMNS}}$ are  given in Eq. \eqref{eq:NOupmns}. 
The experimental constraints of Br($e_b\to e_a \gamma$) and $(g-2)_{e_a}$ are given in Table \ref{t_cabrEx} 
\begin{table}[ht]
	\centering 
	\begin{tabular}{cc}
\hline 
		$\Delta a_{\mu}$:  & 	$|c_{(22)R}|\in \left[1.44, \; 2.14\right]  \times 10^{-9} \; \mathrm{GeV}^{-1}$	\\
$\Delta a_{e}$: &  	$|c_{(11)R}|\in \left[1.73, \;7.57\right]  \times 10^{-11} \; \mathrm{GeV}^{-1}$\\
Br$(\mu\to e\gamma)$: & $|c_{(21)R}|, |c_{(12)R}|< 1.27\times 10^{-12} \; \mathrm{GeV}^{-1}$\\ 

 Br$(\tau\to e\gamma)$: & $|c_{(31)R}|, |c_{(13)R}|<4.15\times 10^{-10} \; \mathrm{GeV}^{-1}$\\ 
 Br$(\tau\to \mu\gamma)$:  & $|c_{(32)R}|, |c_{(23)R}|<4.68\times 10^{-10} \; \mathrm{GeV}^{-1}$\\
 \hline 
	\end{tabular}
	\caption{Allowed values of $c_{(ab)R}$ from experiments} \label{t_cabrEx}
\end{table}
for $c_{(ab)R}$, leading to experimental constraints with the 1$\sigma$ ranges of $\Delta a_{\mu,e}$ given as follows 
\begin{align}
	\label{eq:exprcabr}
\left| \frac{c_{(22)R} }{c_{(32)R} } \right| &> [3.08,\; 4.1],\; \left| \frac{c_{(11)R} }{c_{(31)R}} \right| > 0.06.
\end{align}
Two results in Eqs. \eqref{eq:fR12} and \eqref{eq:exprcabr} predict that the Br$(\tau\to \mu \gamma)$  gives  $\Delta a_{\mu}(H^\pm)\leq1.16\times 10^{-9}$ when $c_{(32)R,0} \left(H^\pm\right)$ is the dominant one-loop contribution; see a numerical illustration  in Fig. \ref{fig_amLFVm}, 
in which  we used the simple example that $\phi=0$ and $M_1=M_2$, $U_N=I_2$. As a result, the future constraint of Br$(\tau \to \mu \gamma)$ will result in a smaller $\Delta a_{\mu}$ than $10^{-9}$.  In contrast, the remaining eight decay rates depend weakly on $\Delta a_{\mu}$. In contrast, the $\Delta a_e$ easily satisfies  the $1\sigma$  range. The allowed points we collected here satisfy all experimental LFV constraints of cLFV, LFV$h$, and LFV$Z$ decays, and the $1\sigma$  range of $\Delta a_e$, while $\Delta a_{\mu} \geq  10^{-10}$. We checked some  different choices of $M_1\neq M_2$ and $c_{\phi}\neq 1$ to confirm that the general results  we indicated here in the 331$\beta$mISS framework are unchanged. In contrast, the LFV decay rates depend weakly on $\Delta a_e$ in the $1\sigma$ range of experiments. In addition,  all the current and future upper constraints always satisfy the $(g-2)_e$ data. Therefore, we conclude that the  mISS mechanism cannot accommodate the recent $1\sigma$ data corresponding  to  $\Delta a_{\mu}>2 \times 10^{-9}$ because of the experimental data of Br$(\tau \to \mu \gamma)$. Similarly,  the IO scheme predicts that

	%

\begin{align}
	\label{eq:fR12IO}
	\left| \frac{c_{(22)R,0} \left(H^\pm\right)}{c_{(32)R,0} \left(H^\pm\right)} \right| (\mathrm{IO})= \left|\frac{\left( U_{\mathrm{PMNS}}\right)_{12} \left( U_{\mathrm{PMNS}}\right)_{21} -\left( U_{\mathrm{PMNS}}\right)_{11} \left( U_{\mathrm{PMNS}}\right)_{22}}{ \left( U_{\mathrm{PMNS}}\right)_{12} \left( U_{\mathrm{PMNS}}\right)_{31} - \left( U_{\mathrm{PMNS}}\right)_{11} \left( U_{\mathrm{PMNS}}\right)_{32}}\right| \simeq 0.854,
\end{align}
implying that  $\Delta a_{\mu}(H^\pm)\leq 5.5\times 10^{-10}$, smaller than the prediction in the NO scheme. 

\begin{figure}[ht]
	%
	\begin{tabular}{ccc}
		\includegraphics[width=5.5cm]{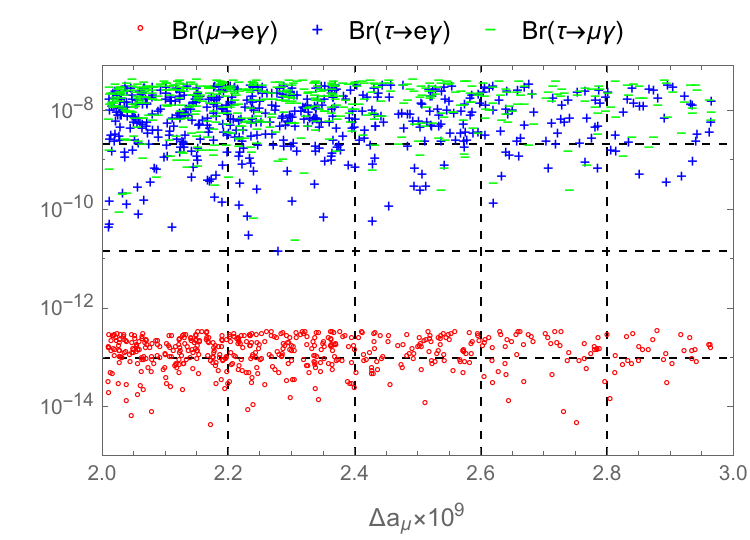}	&  
		\includegraphics[width=5.5cm]{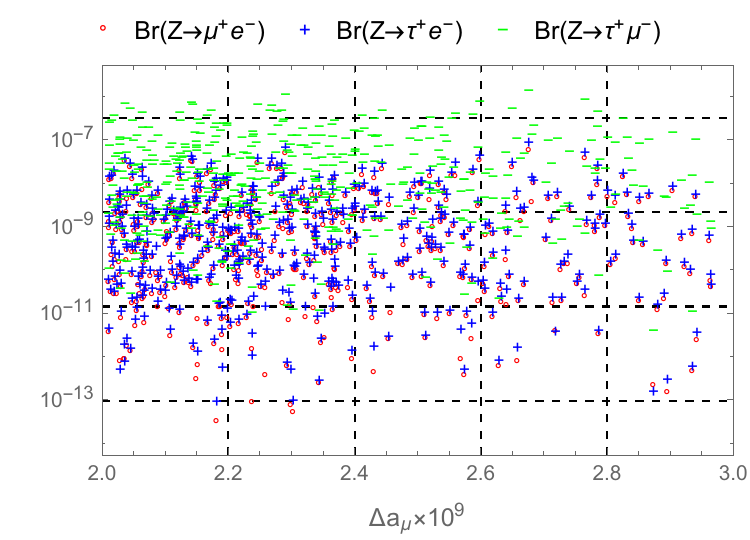}	& 
		\includegraphics[width=5.5cm]{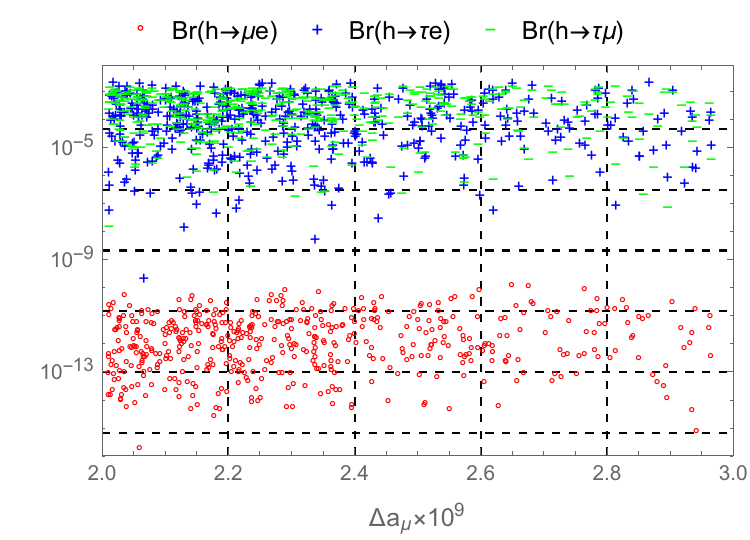}	 
	\end{tabular}
	\caption{ LFV decay rates as functions of $\Delta a_{\mu}$ predicted by the 331$\beta$ISS model (NO). All collected points here satisfy all recent LFV decay rates mentioned in this work.}
	\label{fig_am3LFV}
\end{figure}

\begin{figure}[ht]
	%
	\begin{tabular}{ccc}
		\includegraphics[width=7.5cm]{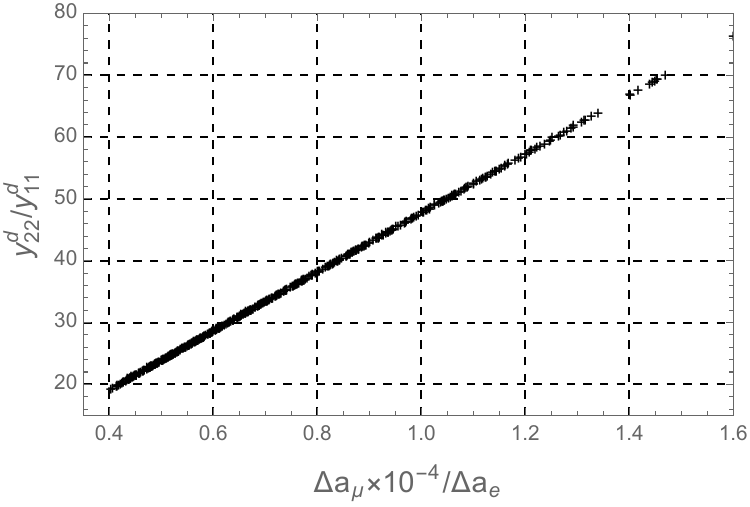}	&  
		\includegraphics[width=7.5cm]{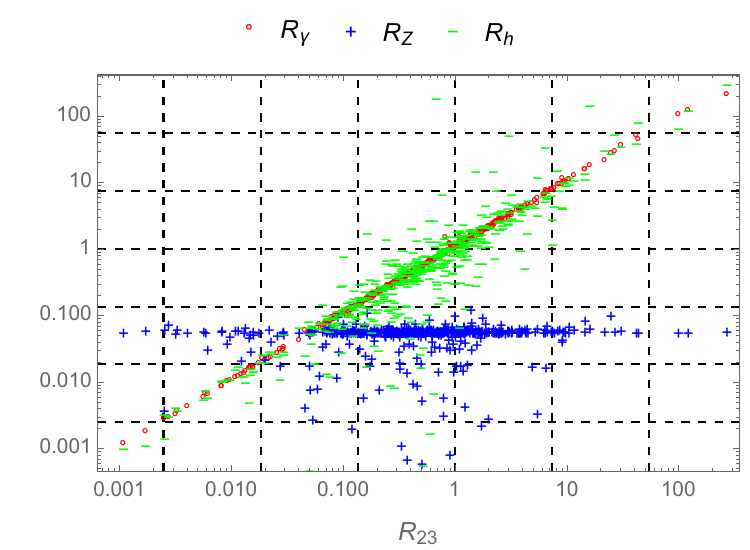}	
		& 
	\end{tabular}
	\caption{ Linear properties of $(g-2)_{e,\mu}$ anomalies (LFV decay rates) in the left (right) panel, showing that all $(g-2)_{\mu,e}$ anomalies and  LFV decays rates get dominant one-loop contributions from Yukawa couplings of singly charged Higgs boson, except the LFV$Z$ ones. }
	\label{f:LcabX}
\end{figure}

Regarding the 331$\beta$ISS model ($K=3$), because the qualitative results from the two NO and  IO are the same, 
the numerical discussions  will focus on only the NO scheme. The constraints of cLFV decays $e_b \to e_a\gamma$ will result in the consequence that  dominant contributions from Eq. \eqref{eq:hatF} will be suppressed with $a\neq b$.  For convenience, all allowed points we collect in the following numerical investigation always satisfy all experimental constraints of  LFV decays under consideration as well as the $1\sigma$ ranges of $\Delta a_{e,\mu}$. Consequently, all the figures we depict in the following discussion will also show the allowed ranges of parameters and relevant experimental constraints.  For simplicity in numerical investigation, we choose here the simplest case in which $M_R=M_0I_3$, and  $\xi=U^e_L=U^e_R=\hat{\mu}_X =I_3$, leading to $\hat{F}_1=\hat{F}_2=\hat{F}_3$. Then, Eq. \eqref{eq:cabRH1} changes into the following simple form:
\begin{equation}
	\label{eq:cabRR1}
c_{(ab)R,0} \left(H^\pm\right) =  \frac{g^2e vt^{-1}_{\beta} x_0^{\frac{1}{2}} c_{\alpha}s_{\alpha}}{32\sqrt{2}  \pi^2 m^2_W }  \left(U_{\mathrm{PMNS}} \hat{x}^{1/2}_{\nu} \hat{\mu}_X^{-1/2}  Y^{h}\right)_{ab} \hat{F}_1.  	
\end{equation}
It was shown that  the diagonal form of $c_{(ab)R,0}(H^\pm)$ given in Eq. \eqref{eq:cabRH1} will result in the allowed range of both $1\sigma$ experimental data of $(g-2)_{e,\mu}$ but  small LFV decay rates satisfying the current experiments \cite{Hue:2021zyw}.  Therefore,  defining the diagonal matrix $Y^d$,  
\begin{align}
\label{eq:defYd}
\left(U_{\mathrm{PMNS}} \hat{x}^{1/2}_{\nu}Y^{h} \right)_{ab}=Y^d_{ab},
\end{align}
then we checked that the diagonal  form of $Y^d_{ab}\equiv Y^d_a \delta_{ab}$ can successfully explain  that both two $\Delta a_{e,\mu}$ and  LFV decay rates  are much smaller than current experimental constraints, except Br$(\mu\to e\gamma)$.   We emphasize that this strict constraint forces $0\simeq  |Y^d_{12}|, |Y^d_{21}|<2\times 10^{-4}$, which we  confirmed by our numerical check.  Therefore, choosing to  scan arbitrary ranges of $Y^h$ to collect allowed points  satisfying   these two relations  is much more difficult than  scanning all entries of $Y^d$ fixing $Y^d_{12}=Y^d_{21}=0$. For this reason, we determine the allowed ranges of the parameter space to collect the allowed points mentioned above; the scanning range of $m_{n_1}$ and $Y^d$ are  chosen as follows:
\begin{align}
\label{eq:Yd}
	m_{n_1}\in  & \left[ 10^{-4},\; 0.032\right]\; \mathrm{eV},\; 
\crn 	Y^d_{22} \in & \left[-5,5\right], \; Y^d_{12}=Y^d_{21}=0,\;  \left| Y^d_{ab}\right| \leq 1 \forall (a,b)\neq \left\{ (12),(21),(22)\right\}. 
\end{align}
All values of $m_{n_1}$ and entries of  $Y^d$ will be checked to satisfy the perturbative limits of the Yukawa couplings $Y^h$ and $Y^{\nu}$ generating $m_D$.  The dependence of LFV decay rates on $\Delta a_\mu$  is depicted in Fig. \ref{fig_am3LFV},
showing  that all LFV decay rates depend weakly of $\Delta a_{\mu}$. The conclusion is the same for $\Delta a_e$. Therefore, the $(g-2)_{\mu,e}$ data cannot predict any significant consequences on LFV decay rates.  The upper bounds of Br$(e_b\to e_a \gamma)$, and Br$(h\to\tau  e_a)$ always reach the current experimental constraints, while those of the remaining are:
\begin{align}
\label{eq:upLFV}
	& \mathrm{Br}(h\to \mu e) \leq  10^{-10} ,\;  \mathrm{Br}(Z\to \mu^+ e^-) \leq 5 \times 10^{-8}, 
\crn & 	\mathrm{Br}(Z\to \tau^+ e^-) \leq 8.3\times 10^{-8},\;  	\mathrm{Br}(Z\to \tau^+ \mu^-) \leq 1.35\times 10^{-6}. 
\end{align}
We can see that  the decay channel $h\to\mu e$ is invisible for near-future experimental searches. These predictions  of Br$(h\to e_be_a)$ distinguish from other current models allowing large values of  max[Br$(h\to \mu e)] \simeq \mathcal{O}(10^{-5})$ \cite{Chen:2023eof}.  There are stricter  constraints on the  parameters than the chosen scanning ranges, $t_{\beta}\leq 14$, $0.12 \leq |s_{2\alpha}|$, $10^{-3}< m_{n_1}<0.03$ eV, $0.02<|Y^d_{11}|<0.2$,  $1<|Y^d_{22}|<4.5$, $|Y^d_{31}|<0.35$, and $\left|m_{H^\pm_1} -m_{H^\pm_2} \right|>216$ GeV, consistent with the dominant contributions shown in Eq. \eqref{eq:cabRH1} which must be non-zero to guarantee large $\Delta a_{e,\mu}$. This allowed range of small $t_\beta$ distinguishes with that requiring very large $t_\beta$ needed in 2HDM models \cite{Iguro:2023tbk}. In addition, the heavy singly charged Higgs masses are allowed, being distinguishable with  many 2HDM models \cite{Ghosh:2023dgk}. Recall that the numerical results we discuss here use the singly charged Higgs couplings given in Eqs. \eqref{eq_lakLR}, in which $	c_{(ab)R,0} \left(H^\pm\right)$ given in Eq. \eqref{eq:cabRH1} is not exactly the same.   But if   $c_{(ab)R,0}\left(H^\pm\right)\varpropto Y^d_{ab}$ give dominant contributions in $\Delta a_{e,\mu}$ and LFV decay rates, the following linear relations will appear: $\Delta a_{\mu}/\Delta a_e \varpropto |Y^d_{22}/Y^d_{11}|$  and $R_{23}\varpropto R_{\gamma}, R_{Z}, R_{h}$, where
$$ R_{23}\equiv \frac{(Y^d_{13})^2 + (Y^d_{31})^2}{(Y^d_{32})^2 + (Y^d_{23})^2},\; R_{\gamma}=\frac{\mathrm{Br}(\tau \to e \gamma)}{\mathrm{Br}(\tau \to \mu \gamma)}, \; R_{Z}=\frac{\mathrm{Br}(Z \to \tau e )}{\mathrm{Br}(Z \to \tau \mu)}, \; R_{h}=\frac{\mathrm{Br}(h \to \tau e )}{\mathrm{Br}(h \to \tau \mu)}.$$
The mentioned relations are illustrated numerically  in Fig. \ref{f:LcabX},
 confirming  that the linear relations happen  for $(g-2)_{e,\mu}$ anomalies,  cLFV, and LFV$h$ decays, but not for LFV$Z$ one.  The reason is that, in the regions of parameters with large LFV$Z$ decay rates, the amplitudes get dominant one-loop contributions from the $W$ exchanges, consisting of the part originating from the massive property of the $Z$, which does not appear for the  photon. The large one-loop contributions from singly charged Higgs result in Br$(h\to \tau e_a)$ being much larger than those predicted by ISS models without Higgs contributions \cite{Arganda:2014dta, Arganda:2016zvc, Thao:2017qtn, Hernandez-Tome:2020lmh}. In conclusion, in the 331$\beta$ISS model, the Yukawa couplings of the singly charged Higgs  are sources of dominant one-loop contributions to $\Delta a_{\mu,e}$,  necessary to explain the $1\sigma$ deviation between the SM and experimental results. These Yukawa couplings will also be important to explain the future experimental results of cLFV and LFV$h$ signals if they are detected.  
 
\begin{figure}[ht]
	%
	\begin{tabular}{ccc}
		\includegraphics[width=5.5cm]{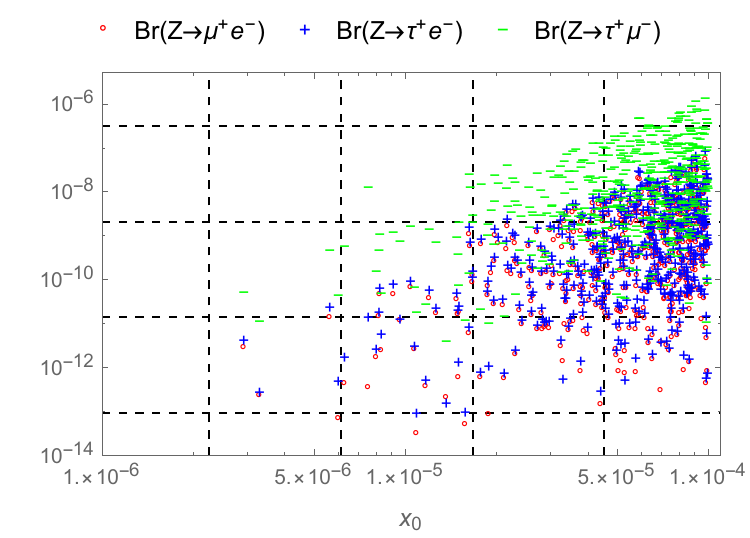}	&  
		\includegraphics[width=5.5cm]{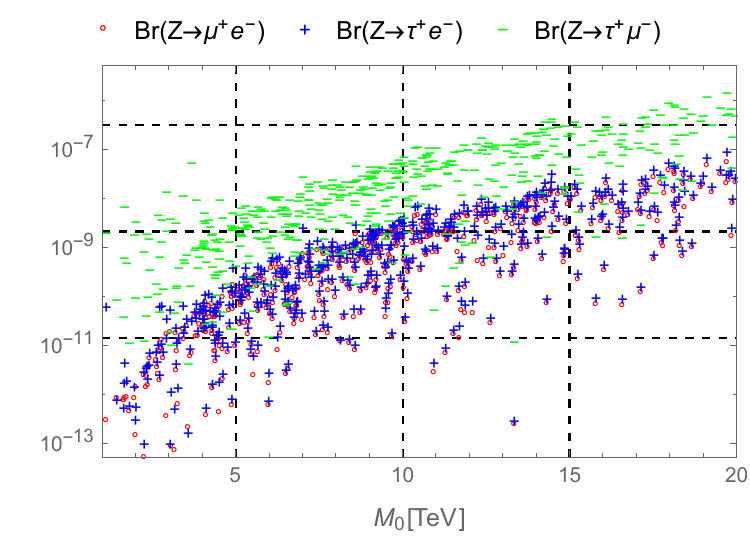}	
		& 
	\includegraphics[width=5.5cm]{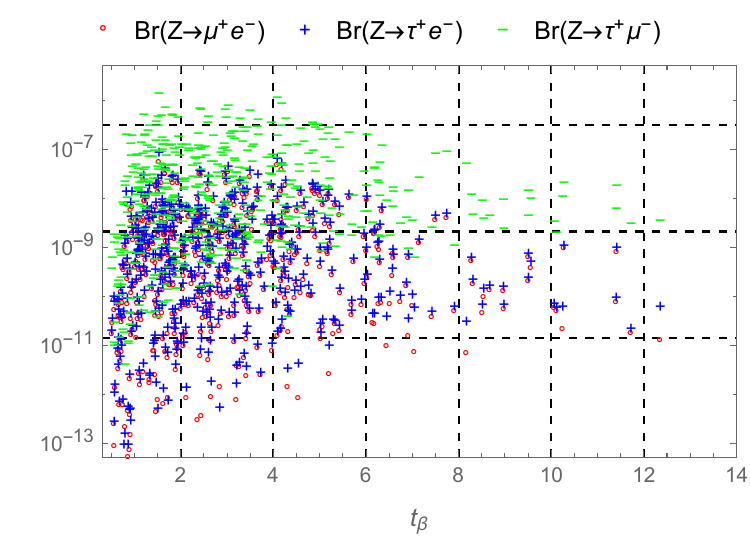}
	\end{tabular}
	\caption{ LFV$Z$ decay rates as  functions of $x_0$, $M_0$, and  $t_{\beta}$ in the 331$\beta$ISS model (NO). The nonunitary factor $x_0$ strongly affects the maximal values of LFV$Z$ decay rates.}
	\label{f:fx-brzc}
\end{figure}

 We can see more interesting properties of LFV$Z$ decays illustrated in Fig. \ref{f:fx-brzc},
where the nonunitary parameter $x_0$ and heavy ISS neutrino mass scale $M_0$ significantly affect  these decay rates,  consistent with ISS models without singly charged Higg exchanges   \cite{Korner:1992an, Abada:2021zcm, Abada:2022asx}.  The future constraint on $x_0$ will result in smaller LFV$Z$ decay rates, giving new indirect constraints along with future direct experimental searches for these channels. Large Br$(Z\to e_b^+e_a)$ prefers moderate values of $t_{\beta}$  \cite{Crivellin:2018mqz},  but smaller than those chosen in  previous discussion of 2HDM models including supersymmetry \cite{Ilakovac:2012sh}. In conclusion, the LFV$Z$ decay rates with ISS neutrinos   are  significantly larger than those predicted by the 2HDM models with standard seesaw neutrinos \cite{Jurciukonis:2021izn}, including the scotogenic  version \cite{Hundi:2022iva}. 

\begin{figure}[ht]
	%
	\begin{tabular}{ccc}
\includegraphics[width=5.cm]{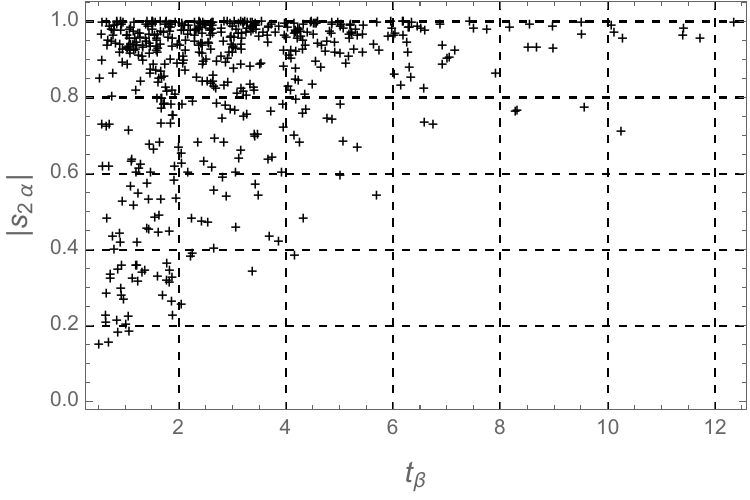}	&  
\includegraphics[width=5.5cm]{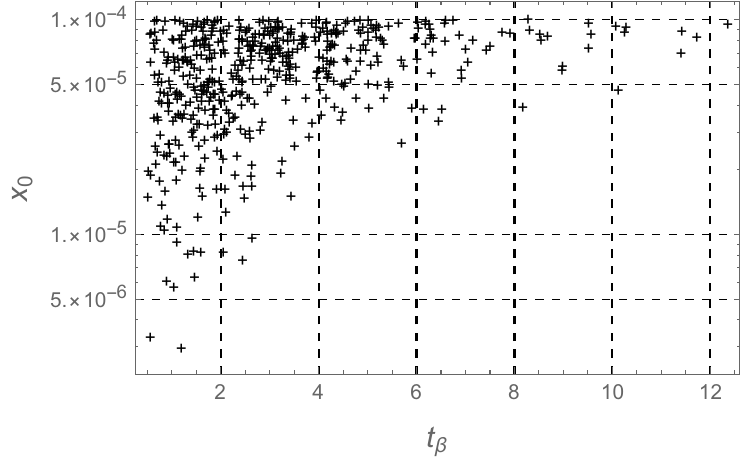}	
& 
\includegraphics[width=5.cm]{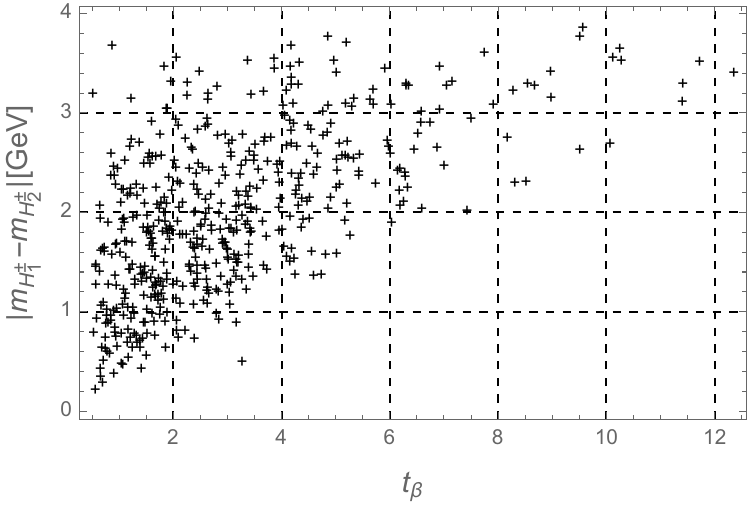}	 \\
\includegraphics[width=5.5cm]{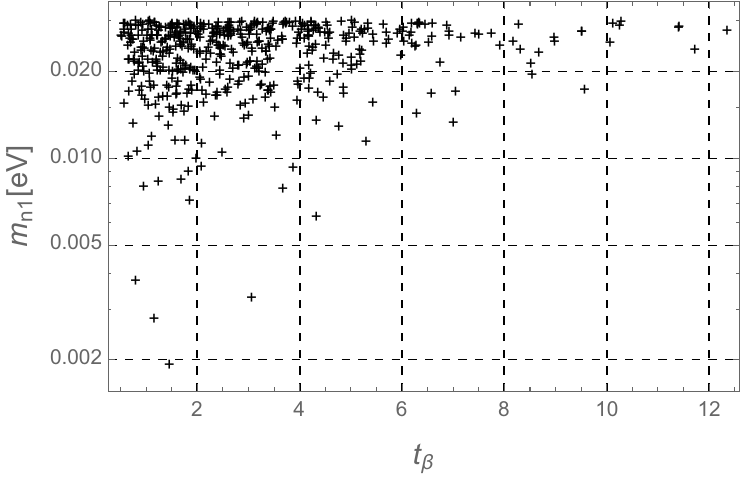}	&  
\includegraphics[width=5.5cm]{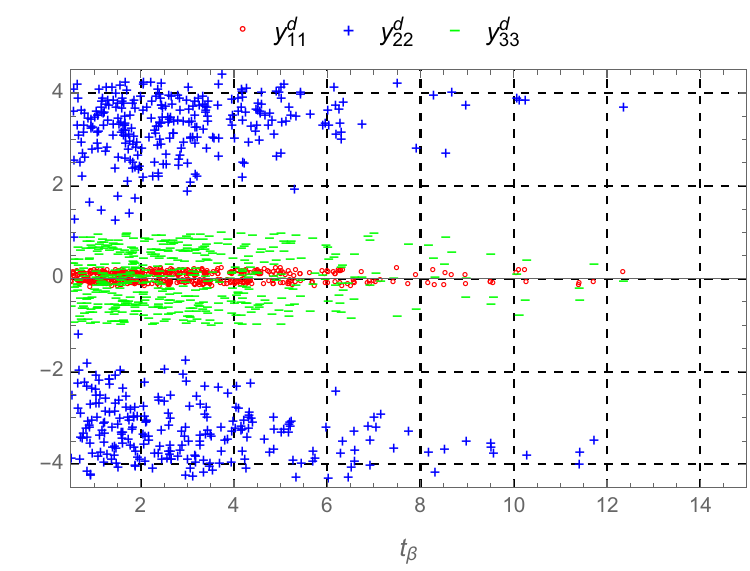}	
& 
\includegraphics[width=5.5cm]{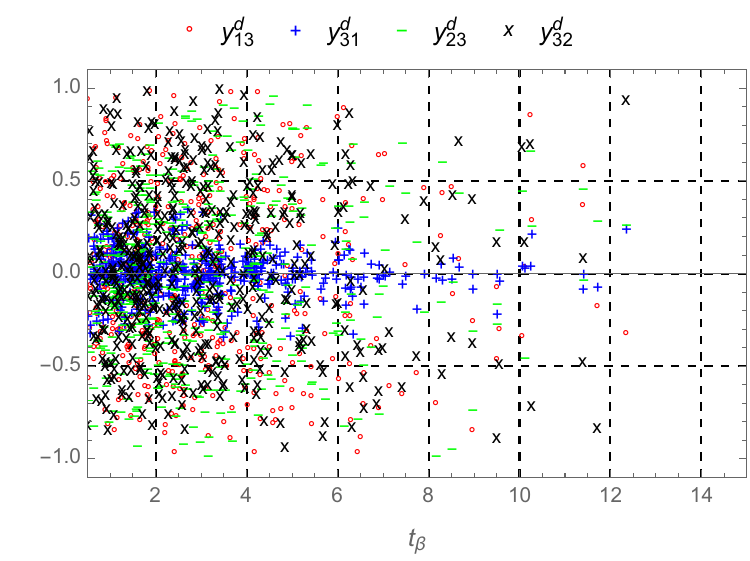}

	\end{tabular}
	\caption{Parameters of the 331$\beta$ISS model as functions of $t_{\beta}$. }
	\label{f:tbX}
\end{figure}

As we discussed above for  the allowed regions of parameter space,  $t_{\beta}$ plays an important role similar to that in the 2HDM models. In the 331$\beta$ model,  the first two components of Higgs triplets $\rho$ and $\eta$ play the same roles discussed in 2HDM \cite{Okada:2016whh, Fan:2022dye, Suarez:2023ozu}. Therefore, the Higgs bosons masses and $t_{\beta}$ are also constrained from experiments relating 2HDM models;  namely,  light $m_{H^\pm_{1,2}}$ requires large $t_{\beta}$. Also, the  popular supersymmetric models inheriting  properties of the 2HDM of type II require   large $t_{\beta}$ to get max[$\Delta a_{\mu}]=10^{-9}$ \cite{Athron:2021iuf, Li:2022zap}. In contrast, the $331\beta$ISS model predicts that  small $t_{\beta}$  are supported  to get large $\Delta a_{e,\mu}$ as well as small $x_0$ consistent with the non-unitary  constraints of the active neutrino mixing matrix consistent with  future experiments.   Figure \ref{f:tbX} shows the dependence of free parameters  on $t_{\beta}$,
pointing out that  small $t_{\beta}$ allows small $x_0$ and heavy $m_{H^\pm_{1,2}}$. Namely, the large values of $t_\beta$ in the allowed regions require large values of many parameters  such as  $|s_{2\alpha}|$, $x_0$, $\left| m_{H^\pm_1} -m_{H^\pm_2} \right|$, $m_{n_1}$, and $Y^d_{22}$, so that the 1$\sigma$ allowed ranges of $\Delta a_{\mu,e}$. On the other hand, small $Y^d_{ab}$ with $a\neq b$ are allowed because they result in small cLFV and LFV$h$ decay rates. 

The allowed regions of parameters shown in Fig. \ref{f:tbX} may affect the new physics signals at LHC relating to singly charged Higgs bosons. Namely, the non-zero couplings  $H^\pm_k n_a e_{a}^\mp$  given in Eq. \eqref{eq_lakLR}  may be large, leading to promising signals of  decays  $H^\pm_k \to e_a^\pm n_a$ searched at LHC such as  $H^\pm \to \tau^\pm \nu_\tau$ \cite{CMS:2019bfg}. Consequently, a lower bound of $m_{H^\pm_k} \geq 500$ GeV is needed. The different values of $t_{\beta}$ also affect the singly charged Higgs couplings with quarks similarly to the 2HDM predictions, because the couplings  $H^\pm_{1,2} \bar{t}b \varpropto \frac{m_t}{vt_{\beta}} \times \{ c_{\alpha}, s_{\alpha} \} $  and $H^\pm_{1,2} \bar{u}d, H^\pm_{1,2} \bar{c}s\varpropto \frac{m_{u,c}t_{\beta}}{v} \times \{ c_{\alpha}, s_{\alpha} \} $ relating to the experimental searches at LHC,  excluding large $t_{\beta}>34$ and $m_{H^\pm_k}>790$ GeV for $H^+\to t\bar{b}$ \cite{ATLAS:2021upq}. In general,  the allowed regions with small $t_{\beta}$ and large $m_{H^\pm_k}$ shown in Fig. \ref{f:tbX} are still favored by LHC searches.  Apart from that, the Yukawa couplings $Y^h_{ab}$ affect significantly the couplings of charged Higgs with heavy neutrinos, implying that the decays $H^\pm \to e^\pm_a n_i$ may appear at LHC \cite{Haba:2011nb, Guo:2017ybk, Tang:2017plx}. In conclusion, the allowed ranges shown in Fig. \ref{f:tbX} will be useful for further studies of these signatures at LHC.

\begin{figure}[ht]
	%
	\begin{tabular}{ccc}
		\includegraphics[width=7.5cm]{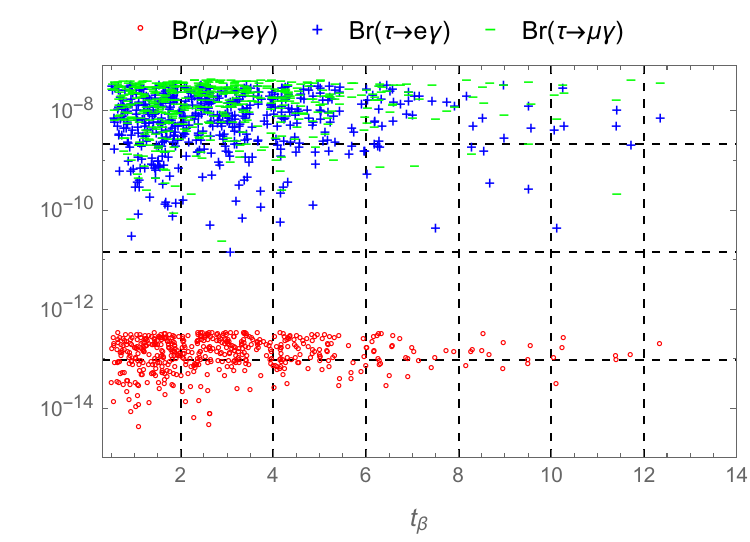}	&  
		& 
		\includegraphics[width=7.5cm]{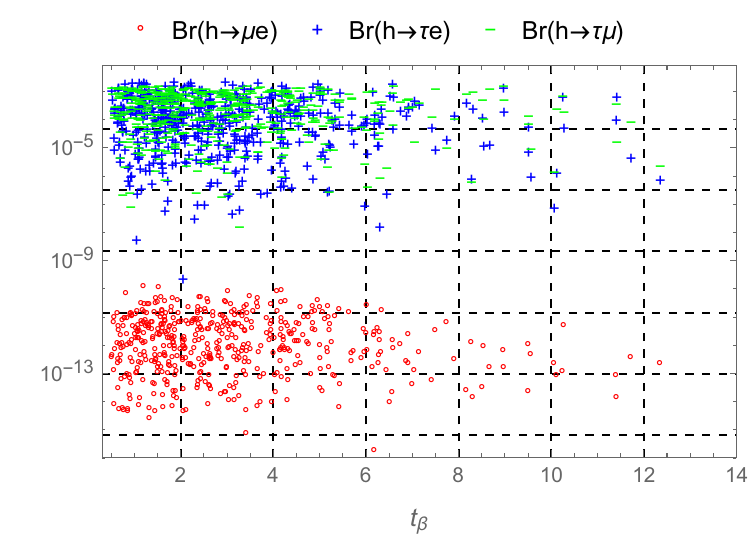}
	\end{tabular}
	\caption{ LFV decay rates as  functions of $t_{\beta}$ in the 331$\beta$ISS model (NO) with $Y^d_{12}=Y^d_{21}=0$, while all remaining entries of $Y^d$ are non-zeros. Large $t_{\beta}$ supports only large values of  Br$(\mu \to e \gamma)$, Br$(\tau \to \mu \gamma)$, and Br$(h\to \tau \mu)$. }
	\label{f:tbLFV}
\end{figure}

Finally, we pay attention to the behavior of cLFV and LFV$h$ decay rates on   $t_{\beta}$ corresponding to some interesting limits of $Y^d$. This may help to understand the properties of LFV decays and the consistency with the new physics at LHC we discussed above.  Illustrations are depicted  in Fig. \ref{f:tbLFV}. 
   Because of very strict experimental constraints on Br$(\mu \to e \gamma)$, the condition $Y^d_{12}=Y^d_{21}=0$ is necessary,  leading to a property shown in the left panel of Fig. \ref{f:tbLFV} that  large $t_{\beta}$ results in large enough Br$(\mu \to e \gamma)$  to reach the current experimental bound. This also gives a consequence of suppressed Br$(h\to \mu e)$.   We note here that large values of Br$(h\to \tau e,\tau \mu)$ up to the incoming experimental are allowed  with small $t_{\beta}$ \cite{Arganda:2015naa}. This property distinguishes  2HDM models supporting large $t_{\beta}$ \cite{Zhang:2021nzv} or  scotogenic models \cite{Hundi:2022iva, Zeleny-Mora:2021tym},  even without discussion of the experimental $(g-2)_{e,\mu}$ data \cite{Brignole:2003iv, Giang:2012vs, Arganda:2015uca, Arroyo-Urena:2023vfh}.   The 331$\beta$ISS model predicts only large Br$(\mu \to e\gamma)$, Br$(\tau \to \mu \gamma)$, and Br$(h\to \tau \mu)$, and hence the future experimental constraints will give more useful information about $t_{\beta}$.
 
\begin{figure}[ht]
 	%
 	\begin{tabular}{ccc}
 		\includegraphics[width=5.5cm]{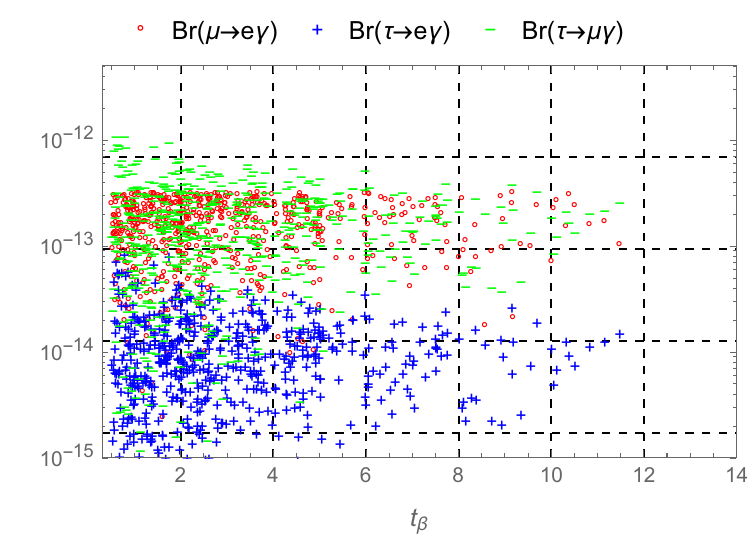}	&  
 		\includegraphics[width=5.5cm]{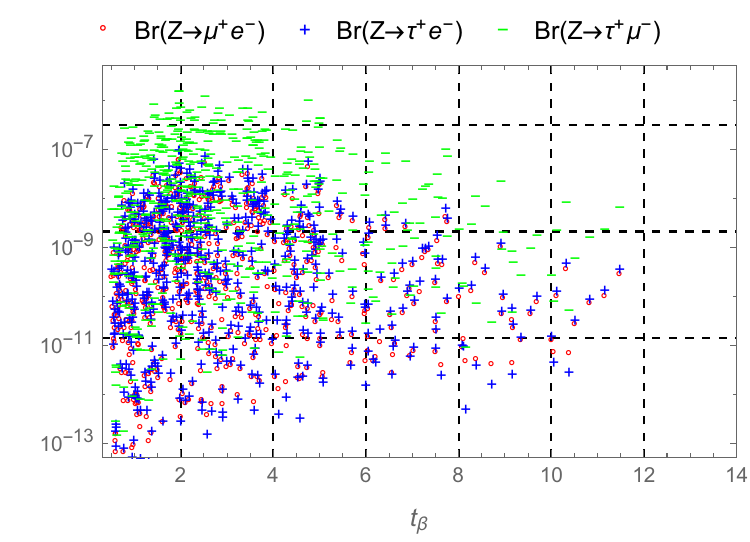}	
 		& 
 		\includegraphics[width=5.5cm]{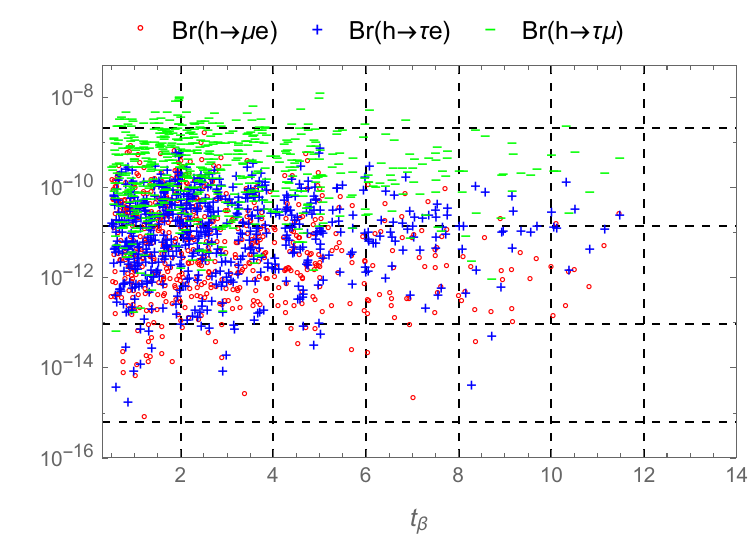}
 	\end{tabular}
 	\caption{LFV  rates vs. $t_{\beta}$ in the 331$\beta$ISS model (NO) with non-zeros $Y^d_{11},Y^d_{22}, Y^d_{33}, Y^d_{12}, Y^d_{21} \neq 0$, while  $Y^d_{13}=Y^d_{31}= Y^d_{23}=Y^d_{32}= 0$. }
 	\label{f:tbLFV0}
 \end{figure}

 Let us discuss on the dependence of  LFV decay rates on $t_{\beta}$ with $Y^d_{ab}\neq 0$ for  $(a\neq b)=\{(12),(21)\}$, while other nondiagonal entries of $Y^d$ vanish, namely $Y^d_{23}= Y^d_{32}=Y^d_{13}=Y^d_{31}=0$. The illustrations are shown in Fig. \ref{f:tbLFV0},
 implying that  all chiral enhancement contributions relating to Yukawa couplings of singly charged Higgs boson  to all LFV decays vanish. We can see that five decay rates including  two cLFV  Br$(\tau \to e\gamma,\mu \gamma)$ and three LFV$h$ ones are invisible for incoming experimental sensitivities, while all LFV$Z$ decays behave  the same as in the cases of $Y^d_{ab}\neq 0$, namely, 
$$ \mathrm{Br}(\tau \to e\gamma,\mu \gamma) \leq \mathcal{O}(10^{-12}), \;  \mathrm{Br}(h \to e_be_a) \leq \mathcal{O}(10^{-8}).$$
Anyway, the above decay rates are much larger than all results indicated in the 2HDM with standard seesaw neutrinos \cite{Zhang:2021nzv, Hundi:2022iva, Zeleny-Mora:2021tym, Jurciukonis:2021izn, Marcano:2019rmk}. In conclusion, the results shown in Fig. \ref{f:tbLFV0} indicate that any signals of Br$(\tau \to e\gamma,\mu \gamma)$ or Br$(h\to \tau \mu, \tau e)$, if observed by incoming experiments, will originate from the respective nondiagonal entries of  $Y^d$ shown in the  right-hand side of  Fig. \ref{f:LcabX}.

\section{\label{sec:con} Conclusions}
 This work discussed the possibilities for the existence of  allowed regions of parameters in two frameworks of  331$\beta$mISS and 331$\beta$ISS models accommodating  $1\sigma$ experimental ranges of $(g-2)_{e,\mu}$ anomalies and all experimental constraints of LFV decays $e_a\to e_a \gamma$, and  $h,Z\to e^\pm_b e^\mp_a$.   Dominant one-loop contributions to $\Delta a_{e_a}$ and  decay amplitudes of cLFV and LFV$h$ appearing in the well-known chiral enhancement parts result in precisely  the 331$\beta$mISS model predicting a strict relation between $\Delta a_{\mu}$  and Br$(\tau \to \mu \gamma)$ because of the strict experimental constraint on Br$(\mu \to e\gamma)$.  This leads to the upper bound $\Delta a_{\mu}\le 1.16\times 10^{-9}$ with  Br$(\tau \to \mu  \gamma)<4.2\times 10^{-8}$.  The 331$\beta$ISS is a more interesting model to accommodate  $1\sigma$ data of $\Delta a_{e,\mu}$ and satisfies all LFV experimental constraints mentioned in this work.   The numerical results and qualitative estimations showed that the allowed regions of parameters successfully explaining  the  $\Delta a_{\mu,e}$ data support the small $t_{\beta} <15$,  heavy singly charged Higgs boson masses at $\mathcal{O}(1)$ TeV scale, and small nonunitary scale $x_0= \mathcal{O}(10^{-6}) $, which will easily satisfy  the future experimental constraints.  In addition, we derived  qualitative relations that $\Delta a_{\mu,e}\varpropto |Y^d_{11,22}|$ and Br$(\tau\to e_a\gamma) , \mathrm{Br}(h\to \tau e_a) \varpropto (|Y^d_{3a}|^2 +Y^d_{a3}|^2 )$ with $e_a=\mu,e$ and $a\neq3$; therefore  $\Delta a_{e_a}$ and LFV decay rates depend weakly on each other, except LFV$Z$ ones. Finally, the allowed regions with small $t_{\beta}$ predicted by the two models under consideration still predict large LFV$h$ and LFV$Z$ decay rates, reaching incoming experimental sensitivities, except the suppressed Br$(h\to \mu e)<10^{-10}$.  The allowed regions with small values of $t_{\beta}$ and heavy singly charged Higgs bosons will be distinguished from those indicated in available models, in which large $t_{\beta}$ is a necessary condition supporting large $\Delta a_{\mu,e}$ and/or LFV decay rates enough to be detected by incoming experiments.

\section*{Acknowledgments}
We thank Doctor Arindam Das, Doctor Xabier Marcano Imaz, and Dr. Gerardo Hernandez Tome for their discussions. This research is funded by Vietnam National
University HoChiMinh City (VNU-HCM) under grant number “C2022-16-06”.  

\appendix
\section{ \label{app_Higgs} Higgs sector}
The scalar potential is \cite{Hue:2021zyw}
\begin{align}
	V_{\mathrm{h}}&=\mu_1^2 \eta^{\dagger}\eta+\mu_2^2\rho^{\dagger}\rho+\mu_3^2\chi^{\dagger}\chi
	+\lambda_1 \left(\eta^{\dagger}\eta\right)^2
	+\lambda_2\left(\rho^{\dagger}\rho\right)^2
	+\lambda_3\left(\chi^{\dagger}\chi\right)^2\crn
	&+ \lambda_{12}(\eta^{\dagger}\eta)(\rho^{\dagger}\rho)
	+\lambda_{13}(\eta^{\dagger}\eta)(\chi^{\dagger}\chi)
	+\lambda_{23}(\rho^{\dagger}\rho)(\chi^{\dagger}\chi)\crn
	&+\tilde{\lambda}_{12} (\eta^{\dagger}\rho)(\rho^{\dagger}\eta) 
	+\tilde{\lambda}_{13} (\eta^{\dagger}\chi)(\chi^{\dagger}\eta)
	+\tilde{\lambda}_{23} (\rho^{\dagger}\chi)(\chi^{\dagger}\rho)  + \sqrt{2} f\left(\epsilon_{ijk}\eta^i\rho^j\chi^k +\mathrm{h.c.} \right) 
	\crn& +\mu_4^2 h^+h^- + f_h \left( \rho^\dagger \eta h^+ +\mathrm{h.c.}\right) + (h^+h^-) \left( \lambda^h_1 \eta^{\dagger}\eta+ \lambda^h_2\rho^{\dagger}\rho +\lambda^h_3 \chi^{\dagger}\chi\right) + \lambda^h_4 \left(h^+h^-\right)^2.  \label{eq_hpo1}	
\end{align}
The neutral components of all Higgs triplets are expanded as follows:
\begin{align}
	\eta^0 =& \frac{1}{\sqrt{2}}\left( v_1 +r_\eta +ia_\eta\right) ,\; \rho^0 = \frac{1}{\sqrt{2}}\left( v_2 +r_\rho +ia_\rho\right) ,\; \chi^0 = \frac{1}{\sqrt{2}}\left( v_3 +r_\chi +ia_\chi\right). 
\end{align}

The relations between the mass and flavor eigenstates of singly charged Higgs bosons are 
\begin{align}
	\begin{pmatrix}
		\eta^{\pm}	&  \\
		\rho^{\pm}	&  \\
		h^{\pm}	& 
	\end{pmatrix} &=\left(
	\begin{array}{ccc}
	c_{\beta } & c_{\alpha } s_{\beta } & s_{\alpha } s_{\beta } \\
	-s_{\beta } & c_{\alpha } c_{\beta } & c_{\beta } s_{\alpha } \\
	0 & -s_{\alpha } & c_{\alpha } \\
	\end{array}
	\right)
	\begin{pmatrix}
		\phi_W^{\pm}	&  \\
		H_1^{\pm}	&  \\
		H_2^{\pm}	& 
	\end{pmatrix}\;  
	\label{scHigg},
\end{align}
where $\phi_W^\pm$ are the Goldstone bosons of $W^\pm$, $H^\pm_{1,2}$ are two physical states  with masses $m_{H^\pm_{1,2}}$, and $\alpha$ is a new mixing parameter. The parameters  $\alpha$, and $m_{H^+_{1,2}}$ are chosen as free ones, while  $\mu_4$, $f$, and $f_h$ are determined as follows:
\begin{align}
	\label{eq_Higgscouplings}
f=& -\frac{c_{\beta } s_{\beta } \left(2 c_{\alpha }^2 m_{H_1^{\pm }}^2 +2 s_{\alpha }^2 m_{H_2^{\pm }}^2 -\tilde{\lambda }_{12} v^2\right)}{2 v_3},
\crn \mu_4^2=& s_{\alpha }^2 m_{H_1^{\pm }}^2 +c_{\alpha }^2 m_{H_2^{\pm }}^2  - \frac{1}{2} \left(\lambda _1^h s_{\beta }^2  v^2 + \lambda _2^h c_{\beta }^2  v^2 +v_3^2 \lambda _3^h\right),
\crn f_h= &-\frac{\sqrt{2} c_{\alpha } s_{\alpha } (m_{H_1^{\pm }}^2-m_{H_2^{\pm }}^2)}{v}. 
\end{align}

The mixing angles of Higgs boson will be defined through the ratios
\begin{equation}\label{eq_sij}
	s_{ab}\equiv \sin\theta_{ab} = \frac{v_a}{\sqrt{v_a^2 +v_b^2}}, \; c_{ab}=\sqrt{1-s^2_{ab}} ,\;t_{ab} =\frac{s_{ab}}{c_{ab}},\; t_{12}= t_\beta^{-1}.
\end{equation}
where $a\neq b$ and $a,b=1,2,3$.

The identification of the SM-like Higgs boson is presented in Ref. \cite{Hung:2019jue}. For neutral Higgs bosons, to avoid the tree-level contribution of SM-like Higgs bosons to the  flavor-changing neutral currents (FCNCs) in the quark sector,   we follow the aligned limit introduced in Ref.~\cite{Okada:2016whh}, namely,
\begin{equation}\label{eq_alignH0}
	f = -\frac{\lambda_{13}v_3}{t_{\beta}}= -\lambda_{23}t_{\beta}v_3.
\end{equation}
From this, we will choose $\lambda_{23}$ and $\lambda_{13}$ as functions of $f$, which is given by Eq. \eqref{eq_Higgscouplings}. As a result, $r_3\equiv h^0_3$ is a physical $CP$-even neutral Higgs boson with mass $m^2_{h^0_3}=2 \lambda _3 v_3^2-\frac{f c_{\beta } s_{\beta } v^2}{v_3}$. There are two other neutral $CP$-even Higgs $h^0_1$ and $h^0_2$, in which  $h^0_1\equiv\, h$ is identified with the SM-like Higgs boson found at LHC.   Denote an independent mixing parameter $\delta$ that defines  $\zeta$ relating two bases of neutral $CP$-even Higgs bosons as follows:   
\begin{align}
	\zeta&\equiv -\theta_{21}  + \delta, \; 
	\tan2\delta \sim\mathcal{ O}\left(\frac{v^2}{v_3^2}\right). \label{eq_mixingh0}
\end{align}
This leads to the following relations: 
\begin{align}
	\left(
	\begin{array}{c}
		r_\eta \\
		r_\rho\\
	\end{array}
	\right)&=\begin{pmatrix}
		c_{\zeta}& s_{\zeta} \\
		-s_{\zeta}& c_{\zeta} 
	\end{pmatrix}\left(
	\begin{array}{c}
		h \\
		h^0_2 \\
	\end{array}
	\right). 
\end{align}
Note that $t_{21}=t_{\beta}>0$  and the couplings of the SM-like Higgs boson are the same as those given in the SM when  $\delta=0$. 

Choose $\delta$ and $ m_h$ as input free parameters; the Higgs self-couplings  are 
\begin{align}
	\label{eq_la13} 
	\lambda_{2} =&\frac{\lambda _1 \left(c_{\beta }^4 c_{\delta }^2+c_{\beta }^2 \left(s_{\beta }^2 +2 s_{\delta }^2\right)-2 c_{\beta } s_{\delta } s_{\zeta }+s_{\beta }^4 s_{\delta}^2\right)}{t_{\beta}^2 s_{\zeta }^2}  
\crn &+ \frac{m_h^2 \left(c_{\beta }^2 \left(s_{\delta }^2-c_{\delta }^2\right)-4 c_{\beta } c_{\delta } s_{\beta } s_{\delta }+s_{\beta }^2 \left(c_{\delta }^2-s_{\delta }^2\right)\right)}{2 s_{\beta }^2 v^2
	s^2_{\zeta}} -\frac{f s_{\delta } v_3 (s_{\delta }-2 c_{\beta } s_{\zeta })}{2 c_{\beta } s_{\beta }^3 s_{\zeta }^2 v^2}, \crn 
	\lambda_{12} =& \frac{2 \lambda _1}{t_{\beta } t_{\zeta }} -\frac{m_h^2}{c_{\beta } s_{\beta } t_{\zeta } v^2} -\frac{f s_{\delta } v_3}{c_{\beta }^2 s_{\beta } s_{\zeta } v^2}. 
\end{align}

The  Higgs self-couplings  satisfy all constraints discussed recently to guarantee the vacuum stability of the Higgs potential, the perturbative limits, and the positive squared masses of all Higgs bosons \cite{Costantini:2020xrn, Li:2021poy, Cherchiglia:2022zfy}. We note that  in the case of absence of the relations in Eq.~\eqref{eq_alignH0}, the mixing between SM-like Higgs bosons with other heavy neutral Higgs is still suppressed due to large $v_3>5$ TeV enoungh to cancel the FCNCs  in 3-3-1 models~\cite{Huitu:2019kbm}. We list here the necessary conditions relating to the involved Higgs  couplings:
\begin{align}
\label{eq:Higgscouplingc}
	 & \lambda_1,\lambda_2>0;\; \lambda_{12} +2\sqrt{\lambda_1\lambda_2}>0,\; \lambda_{12}+\tilde{\lambda}_{12} +2\sqrt{\lambda_1\lambda_2}>0, \;
	|\lambda_{ij}| <4\pi.
\end{align} 
 For simplicity, we will fix $\lambda^h_1=\lambda^h_2=0$ in numerical investigation.

\section{ \label{app:ebagaloop} Master one-loop functions to $(g-2)_{e_a}$ anomalies and decays $e_b\to e_a \gamma$   }

 The relevant functions for one-loop contributions to cLFV decays and $(g-2)_{e_a}$ anomalies mentioned in this work are \cite{Crivellin:2018qmi}
\begin{align}
	\label{eq_fPhiV}	
	&f_\Phi(x)=2\tilde{g}_\Phi(x)=\dfrac{x^2-1-2x\log x}{4(x-1)^3},
	\nonumber \\
		&\tilde{f}_{\Phi}(x)=\dfrac{2x^3+3x^2-6x+1-6x^2\log x}{24(x-1)^4},\non\\
	&\tilde{f}_{V}(x)=\dfrac{-4x^4+49x^3-78x^2+43x-10-18x^3 \ln x}{24(x-1)^4}, 
\end{align}
which consistent with previous work for 3-3-1 models \cite{Hue:2017lak, Lavoura:2003xp}.  

\section{\label{app_Zeba}  One-loop contribution to $Z\to e_b^+ e_a^-$}
\subsection{PV-functions and general analytic formulas}
The one-loop contributions here are calculated using the  notations of Passarino functions \cite{tHooft:1972tcz, Denner:2005nn} given in Ref.~\cite{Nguyen:2020ehj}, consistent with LoopTools \cite{Hahn:1998yk};  see a detailed discussion in Ref. \cite{Hue:2015fbb}.  The  Passarino-Veltman (PV) functions used in this work are defined as follows: $B^{(i)}_{\mu}\equiv B_1^{(i)}\times (-1)^i p_{i\mu}$ with $i=1,2$;  $C_{\mu}\equiv  - C_1 p_{1\mu} +C_2 p_{2\mu}$; and 
$$ C_{\mu\nu}\equiv C_{00} g_{\mu \nu}+C_{11} p_{1\mu} p_{1\nu} + C_{22} p_{2\mu} p_{2\nu} -C_{12}(p_{1\mu} p_{2\nu} +p_{2\mu} p_{1\nu}).$$   Particular notations of the PV-functions used  in our formulas are: $B^{(i)}_{0,1}=B_{0,1}(p_i^2; M^2_0,M^2_i)$, $C_{0,1,2}=C_{0,1,2}(p_1^2,q^2, p_2^2; M_0^2,M_1^2,M_2^2)$, and  $B^{(12)}_0( M^2_1,M^2_2)=B_0(q^2; M^2_1,M^2_2)$. The LFV decays of $B=h,Z$ will give $q^2 = (p_1+p_2)^2=m_{B}^2$, while  the external momenta are always fixed as  $p_1^2=m^2_{a}$, $p_2^2=m^2_{b}$. Linear combinations of PV-functions used in this work are 
\begin{align}
	\label{eq_Xidef}
	X_0&\equiv C_0 +C_1 +C_2, \; 
	   X_1\equiv C_{11} +C_{12} +C_1, \;
   X_2 \equiv C_{12} +C_{22} +C_2, \; 
   X_3\equiv C_1 +C_2,
	\crn   X_{012}&\equiv X_0+X_1 +X_2, \; X_{ij}= X_i+X_j.
\end{align} 

We  summarize here the Lagrangian parts appearing 331$\beta$ model that give one-loop contributions to all relevant LFV decays. The details are mentioned in Ref. \cite{Hung:2019jue}. 

The couplings of gauge bosons $G_{\mu}$ including $Z$, photon $A_{\mu}$, $W^\pm$, and $Y^{\pm Q_A}$ with leptons  arise from their covariant kinetic parts
\begin{align}\label{eq_Lkinf}
	\mathcal{L}^{f}_{\mathrm{kin}} =& \sum_{a=1}^3\left( \overline{L_{aL}}\gamma^{\mu}D_{\mu}L_{aL} + \overline{\nu_{aR}}\gamma^{\mu}\partial_{\mu}\nu_{aR} + \overline{e_{aR}}\gamma^{\mu} D_{\mu}e_{aR} + \overline{E_{aR}}\gamma^{\mu} D_{\mu}E_{aR}\right)
	\crn
=& \sum_{F_1,F_2} \left[ \overline{F_1}\gamma^{\mu} e\left( g^L_{ZF_{12}} P_L +g^R_{ZF_{12}} P_R\right) F_2Z_{\mu}  \right] + \sum_{F} e Q_F \overline{F}\gamma^{\mu}F A_{\mu}
	\crn&+  \sum_{F,a} \left[g^L_{aFG}\overline{F}\gamma^{\mu}P_Le_a G^{+Q_G}_{\mu} + \mathrm{h.c.}\right] +\dots,
\end{align}
where new fermions $F$ and  $F_{1,2}$ run over all relevant leptons   appearing in the 331$\beta$ISS$(K,K)$ model and $Q_F$ is the electric charge of the fermion $F$. The electric charge  conversation results in  $Q_G=Q_F+1$. The first term in the second line of Lagrangian \eqref{eq_Lkinf} is valid for only Dirac fermions. The general Feynman rules for vertex and couplings corresponding to the Lagrangian \eqref{eq_Lkinf} are listed in the  lines  2, 4, and 6 of the  Table \ref{tab_FeynruleZeba}.
\begin{table}[ht]
	\centering 
	\begin{tabular}{|c|c|c|c|}
		\hline
		Vertex & Coupling & Vertex & Coupling \\
		\hline
		$\overline{F}e_{a}H^{+Q}$ & $-i \left( g^{R}_{aFH} P_R +g^{L}_{aFH}P_L\right) $ &
		$\overline{e_a}F H^{-Q}$ & $ -i \left( g^{R*}_{aFH} P_L  +g^{L*}_{aFH}P_R\right) $ \\ 
		\hline
		$\overline{F} e_{a}G^{+Q}_{\mu}$ &$i g^{L}_{aFG}\gamma^\mu P_L $ &$\overline{e}_a F G^{-Q}_\mu$
		& $i g^{L*}_{aFG}\gamma^\mu P_L $\\
		\hline
		$A^{\la}G^{+Q\mu}G^{-Q\nu}$&$-ieQ\Gamma_{\la \mu \nu}(p_0,p_{+},p_{-}) $&$A^{\mu}H^{+Q}H^{-Q}$ & $ieQ(p_{+} -p_{-})_\mu$\\
		\hline
		$A^{\mu}\bar{e}_ae_a$&$-ie\gamma_\mu $&$A^{\mu}\overline{F} F$&$ieQ_F \gamma_\mu  $\\
		\hline
		$Z^{\la}G^{+Q\mu} G^{-Q\nu}$&$-i eg_{ZGG}\Gamma_{\la \mu \nu}(p_0,p_{+},p_{-}) $&$Z^{\mu}H_1^{+Q}H_2^{-Q}$ & $i eg_{ZH_1H_2}(p_{+} -p_{-})_\mu$\\
		\hline
		$Z^{\mu}\bar{e}_ae_a$&$ie \gamma_\mu \left( g^L_{Ze}P_L + g^R_{Ze}P_R\right)$ & $Z^{\mu}\overline{F_1} F_2$& $i e\gamma_\mu \left( g^L_{ZF_{12}}P_L + g^R_{ZF_{12}}P_R\right)$\\
		\hline
		$Z^{\mu}G^{+Q\nu}H^{-Q}$&$ i eg_{ZGH}g_{\mu \nu} $& $Z^{\mu}G^{-Q\nu} H^{+Q}$&$i eg^*_{ZGH}g_{\mu \nu} $ \\
		\hline
		$h G^{+Q\mu}G^{-Q\nu}$& $ig_{hGG} g_{\mu\nu}$ &$h H^{+Q}_1H^{-Q}_2$ & $i \lambda_{hH_1H_2}$\\
		\hline
		$h\bar{e}_ae_a$& $-i \frac{m_{a}}{v} \delta_{hee}$ & $h\overline{F_1} F_2$& $-i  \left( g^L_{hF_{12}}P_L + g^R_{hF_{12}}P_R\right)$\\
		\hline
		$hH^{+Q}G^{-Q\mu}$& $ -i g^*_{hHG}(p_0 -p_+ )_{\mu}$ & $hH^{-Q} G^{+Q\mu} $&   $ i g_{hHG}(p_0 -p_- )_{\mu}$\\
		\hline
	\end{tabular}
	\caption{General Feynman rules for vertices and couplings   giving one-loop contributions  to LFV decays and $(g-2)_{e_a}$ anomalies in the 331$\beta$ISS$(K,K)$ model. 
		\label{tab_FeynruleZeba}}
\end{table}
The remaining vertices and couplings in this Table are defined in  the following. 

The Yukawa Lagrangian \eqref{eq_ylepton1} results in the couplings of Higgs bosons and leptons. The couplings of the SM-like Higgs boson $h$ and charged Higgs relevant with this work have the following forms: $ \mathcal{L}^Y= - h \bar{F}_1 \left( g^L_{hF_{12}}P_L +g^R_{hF_{12}}P_R\right)F_2 -  \bar{F}\left( g^R_{aFH}P_R +g^L_{aFH}P_L \right)e_a H^{+Q} +\mathrm{h.c.} $. The corresponding Feynman rules are shown in Table \ref{tab_FeynruleZeba}. Note that the couplings relating to  a SM charged lepton $e_a$ must be close to those of the SM. 

The couplings of Higgs and gauge bosons are contained in the covariant kinetic terms of the Higgs bosons
\begin{align}
	\label{eq_lkHiggs}
	\mathcal{L}^{S}_{\mathrm{kin}}=& \sum_{S=\chi, \rho, \eta}\left(D_{\mu}S \right)^{\dagger} \left(D^{\mu}S \right) \crn
	= & \sum_{G}g_{hGG} g_{\mu\nu}hG^{-Q\mu} G^{Q\nu} \crn 
	& + \sum_{H,G}\left[ -ig^*_{hHG} \left( H^{+Q}\partial_{\mu}h -h\partial_{\mu}H^{+Q} \right) G^{-Q\mu} + ig_{hHG} \left( H^{-Q}\partial_{\mu}h -h\partial_{\mu}H^{-Q} \right)G^{Q\mu} \right]\crn 
	& +\sum_{H} ie Q A^{\mu}\left( H^{-Q}\partial_{\mu} H^{Q} -H^{Q}\partial_{\mu} H^{-Q} \right) + \sum_{k,l}ieg_{ZH_kH_l}Z^{\mu} \left( H_k^{-Q}\partial_{\mu}H_l^{Q} -H^{Q_k }\partial_{\mu}H_l^{-Q} \right)  
	\crn &
	+\sum_{H,G} Z^{\mu}e \left[ ig_{ZGH}  G^{Q\nu}H^{-Q} g_{\mu\nu} + ig^*_{ZGH} G^{-Q\nu}H^{Q} g_{\mu\nu}\right]    +..., 
\end{align}
where sums are taken over  $H,H_k,H_l=H^{\pm},H^{\pm A},H^{\pm B}$,  and gauge bosons $G=W,Y,V$.

The triple couplings of three gauge bosons arise from the covariant kinetic Lagrangian of the non-Abelian gauge bosons
\begin{equation}\label{eq_LDg}
	\mathcal{L}^g_D= -\frac{1}{4}\sum_{a=1}^8F^a_{\mu\nu}F^{a\mu\nu},
\end{equation} 
where 
\begin{equation}\label{eq_Famunu}
	F^a_{\mu\nu}=\partial_{\mu}W^a_{\nu} -\partial_{\nu}W^a_{\mu} + g \sum_{b,c=1}^8f^{abc}W^b_{\mu} W^c_{\nu},
\end{equation}
where $f^{abc}$ $(a,b,c=1,2,...,8)$ are structure constants of the $SU(3)$ group. Changing into the momentum space  using $\partial_{\mu}\to -i p_{\mu}$, the  relevant couplings are  defined as 
\begin{align}\label{eq_ZAgg}
	\mathcal{L}^g_D \rightarrow& -eg_{ZVV} Z^{\mu}(p_0)V^{+Q\nu}(p_+)V^{-Q\lambda}(p_-)\times \Gamma_{\mu\nu\lambda}(p_0,p_+,p_-), \crn 
	&-eQA^{\mu}(p_0) V^{+Q\nu}(p_+)V^{-Q\lambda}(p_-)\times \Gamma_{\mu\nu\lambda}(p_0,p_+,p_-),
\end{align}
where $\Gamma_{\mu\nu\lambda}(p_0,p_+,p_-) \equiv g_{\mu\nu}(p_0-p_+)_{\lambda} +g_{\nu\lambda}(p_+ -p_-)_{\mu} +g_{\lambda\mu}(p_- -p_0)_{\lambda}$, and $G=W,V,Y$. 
This class of couplings was given in Tables \ref{t:hl331iSS} and \ref{table_3gaugcoupling}, for the decays $e_b \to e_a \gamma$ and $Z\to e_a e_b$,  respectively.

One-loop form factors from diagram 1 of Fig. \ref{fig_LFVZ} calculated in the unitary gauge are well known from Refs. \cite{Abada:2022asx, Hong:2023rhg}, but for only $W$ exchange.  Because the similar diagrams with $V$ exchanges have the same forms of Feynman rules, they can be derived in the same way. The general Feynman rules are listed in Table \ref{tab_FeynruleZeba}, where we use the   relations
 $$\sum_Fg^{L*}_{aFG}g^{L}_{bFG},  \sum_{F_{1,2}}  g^{L*}_{aF_1G} g^{L}_{bF_2G} g^L_{ZF_{12}},  \sum_{F_{1,2}}  g^{L*}_{aF_1G} g^{L}_{bF_2G} g^R_{ZF_{12}}\varpropto \delta_{ab}$$ 
 for both $F=E_a,n_i$ in this work. Therefore, we can write the following expressions for both $W,V\equiv G$: 
\begin{align}
	\bar{a}^{FGG}_l	=& \sum_{F} g^{LL}_{ab} g_{ZGG}
	\left\{ \left[ -4+  \dfrac{m^2_F}{m^2_G} \left(\frac{m_Z^2}{m_G^2} -2\right)\right] C_{00}  +2\left(m^2_Z- m^2_a -m^2_b\right)X_3 
	\right.\crn &\left. \hspace{2.8cm}
	-\frac{1}{m_G^2}  \left[\frac{}{} m^2_Z  \left(2m^2_FC_0 +m^2_aC_1 +m^2_bC_2\right)
	\right.\right.\crn	& \left.\left. \hspace{4cm} 
	-m^2_F\left(B_0^{(1)} +B_0^{(2)}\right) -m^2_aB_1^{(1)} -m^2_bB_1^{(2)}   \right]  \right\} , \label{eq_al1u} 
	%
	%
	\\	\bar{a}^{FGG}_ r=&  \sum_{F} g^{LL}_{ab} g_{ZGG}m_am_b \left[  \left(-4 + \frac{ m^2_Z}{m_G^2} \right) X_3 
	+ \dfrac{m^2_Z -2 m_G^2}{m^4_G}C_{00} \right], \label{eq_ar1u}
	%
	\\	\bar{b}^{FGG}_l= & \sum_{F} g^{LL}_{ab}  g_{ZGG}m_a \left[ 4 \left( X_3 -X_1\right) 	+\dfrac{m^2_Z -2m_G^2}{m^4_G}   \left( m^2_FX_{01} +m^2_bX_2 \right)   -\frac{2m_Z^2}{m_G^2} C_2  
	\right], \label{eq_bl1u} 
	%
	\\	\bar{b}^{FGG}_r =& \sum_{F} g^{LL}_{ab} g_{ZGG}m_b \left[4\left(X_3   - X_2\right)  +\dfrac{m^2_Z -2 m_G^2}{m^4_G}\left( m^2_FX_{02} +m^2_aX_1\right) -\frac{2m^2_Z}{m_G^2} C_1 \right], \label{eq_br1u}
\end{align}
where  $g^{LL}_{ab} \equiv g^{L*}_{aFG} g^{L}_{bFG}$,  $B^{(k)}_{0,1}=B^{(k)}_{0,1}(p_k^2; m_{F}^2,m_G^2)$,  $C_{k,kl}=C_{k,kl}(m_a^2,m_Z^2,m_b^2; m_{F}^2, m_G^2,m_G^2),$ 
and  $X_{0,k,kl}$  are defined in terms of the PV-functions in Eq. \eqref{eq_Xidef} for all $k,l=0, 1,2$. For simplicity, the arguments of  PV-functions and their linear combinations are written specifically  as  $(m_a^2,m_Z^2,m_b^2;m_{F}^2, m_G^2,m_G^2)$ or $(M_0^2,M_1^2,M_2^2)=(m_{F}^2, m_G^2,m_G^2)$ when $q^2$ is determined precisely. We will use this simple notations from now on. 

One-loop form factors from diagram 5    in Fig. \ref{fig_LFVZ} are  
\begin{align}
	\bar{a}^{GFF}_l =& \sum_{F_{1,2}} \frac{g^{LL}_{ab}}{m_G^2} \left\lbrace g^L_{ZF_{12}} \left[  m^2_G \Big(4C_{00}  +2 m^2_aX_{01} +2m^2_bX_{02} -2m^2_Z \left( C_{12}+X_0\right)\Big) \right.\right.\crn
	&\left.\left. \hspace{2.6cm} -\left(m_{F_1}^2 -m^2_a\right)B_0^{(1)} -\left(m_{F_2}^2-m^2_b\right)B_0^{(2)}  +m^2_aB_1^{(1)} +m^2_bB_1^{(2)}
	\right.\right.\crn
	&\left.\left. \hspace{2.6cm}+ \left(m_{F_2}^2m_{a}^2 +m_{F_1}^2m_{b}^2-m_a^2m_b^2 \right) X_0  -m_{F_1}^2m_{F_2}^2C_0 
	%
	-m_{F_1}^2m_{b}^2 C_1 -m_{F_2}^2m_{a}^2 C_2  \frac{}{}\right] \right.\crn
	&\left. \hspace{1.5cm}-g^R_{ZF_{12}} m_{F_1}m_{F_2}\bigg[2m^2_WC_0 -2C_{00} -m^2_aC_{11} -m^2_bC_{22} 
	%
	+\left(m^2_Z -m^2_a -m^2_b\right)C_{12}\bigg]\right\rbrace,  \label{eq_al2u} 
	%
	\\ \bar{a}^{GFF}_r  =&  
	\sum_{F_{1,2}}  \frac{g^{LL}_{ab}m_am_b}{m_G^2} g^L_{ZF_{12}} \left[\frac{}{} 2C_{00}+ 2m_{G}^2X_0 +m^2_aX_1 +m^2_bX_2 
	%
		 -m_Z^2 C_{12}  -m_{F_1}^2C_1 -m_{F_2}^2C_2\right], \label{eq_ar2u}
	\\ \bar{b}^{GFF}_l  =&\sum_{F_{1,2}} \frac{2m_a}{m_G^2}  g^{LL}_{ab} \left[ g^L_{ZF_{12}} \left(-2m_{G}^2 X_{01}   -m^2_bX_2 +m_{F_2}^2C_2\right) -g^R_{ZF_{12}} m_{F_1}m_{F_2}(X_1 -C_1) \right],   \label{eq_bl2u} 
	%
	\\ \bar{b}^{GFF}_r =& \sum_{F_{1,2}} \frac{2 m_b}{m_G^2} g^{LL}_{ab} \Big[g^L_{ZF_{12}} \left( -2m_{G}^2X_{02}  -m^2_aX_1  +m_{F_1}^2C_1\right) - g^R_{ZF_{12}}m_{F_1}m_{F_2}(X_2 -C_2)
	\Big], \label{eq_br2u}
\end{align} 
where  $g^{LL}_{ab}\equiv g^{L*}_{aF_1G} g^{L }_{bF_2G}$, and arguments for PV-funtions are $( m_G^2,m_{F_1}^2, m_{F_2}^2)$.

The   form factors for sum  contributions  from two diagrams 7 and 8 in Fig. \ref{fig_LFVZ} are 
\begin{align}
	\bar{a}^{FG}_l =&   \sum_{F}\dfrac{ g^{L*}_{aFG}g^{L}_{bFG} g^L_{Ze}}{m^2_G(m^2_a -m^2_b)}
	\left\lbrace 2m^2_{F}(m^2_a B_0^{(1)}- m^2_b B_0^{(2)})  + m^4_a B_1^{(1)}  -m^4_b B_1^{(2)}
	\right.\crn&\left.\hspace{5.5cm}+  \left(2 m^2_G +m^2_{F} \right) \left(m^2_aB_1^{(1)} -m^2_bB_1^{(2)} \right)  \right\rbrace. \label{eq_al3u} 
	%
	\\ \bar{a}^{FG}_ r =& \sum_{F}\dfrac{ m_am_b g^{L*}_{aFG}g^{L}_{bFG} g^R_{Ze}}{m^2_G (m^2_a -m^2_b)}   \left\lbrace 2m^2_F(B_0^{(1)}- B_0^{(2)})  + m^2_aB_1^{(1)}  -m^2_b B_1^{(2)}
	\right.\crn&\left. \hspace{5.5cm} +  \left(2m^2_G +m^2_{F}\right) \left(B_1^{(1)} -B_1^{(2)} \right)  \right\rbrace , \label{eq_ar3u}
	\\ \bar{b}^{FG}_{l} =&    \bar{b}^{FG}_{r} = 0, \label{eq_blr3u}
\end{align}
where  $g^{LL}_{ab} \equiv g^{L*}_{aFG} g^{L}_{bFG}$ and $B^{(k)}_{0,1}=B^{(1)}_{0,1}(p_k^2; m_{F}^2,m_G^2)$ with $k=1,2$. 

The two diagrams 2 and 3 appearing in the model under consideration were not discussed previously; we list here for completeness the form factors of diagram 3
\begin{align}
\label{eq:abFVH}
	\overline{a}^{FGH}_l &= \sum_{F} -\frac{g_{ZGH}  g^{L*}_{aFG}}{m_G^2} \left[ g^{L}_{bFH}m_{F}\left(m_G^2C_0  -C_{00}\right)  - g^{R}_{bFH}m_bm_G^2C_2\right], 
\crn 	\overline{a}^{FGH}_r&= \sum_{F} -\frac{g_{ZGH}  g^{L*}_{aFG}}{m_G^2}\times g^{R}_{bFH}m_a\left(C_{00} +m_G^2C_1 \right), 
\crn 	\overline{b}^{FGH}_l &=\sum_{F} -\frac{g_{ZGH}  g^{L*}_{aFG}}{m_G^2} \left[ m_a \left(g^{R}_{bFH} m_bX_2 - g^{L}_{bFH} m_{F}X_{01}  \right)\right], 
\crn	\overline{b}^{FGH}_r &=\sum_{F} -\frac{g_{ZGH}  g^{L*}_{aFG}}{m_G^2} \left[ g^{R}_{bFH} \left(m^2_{F}X_0 +m^2_a X_1 -2m_G^2 C_1\right) -g^{L}_{bFH} m_{F}m_bX_2 \frac{}{}\right],
\end{align}
where arguments for PV-funtions are $(m_{F}^2, m_G^2, m_{H}^2)$.  

The form factors  relating to diagram 3 are 
\begin{align}
	\label{eq:abFVH}
	\overline{a}^{FHG}_l &=\sum_{F} -\frac{g_{ZGH}  g^{L}_{bFG}}{m_G^2} \left[g^{L*}_{aFH} m_F\left( m^2_G C_0 -C_{00}\right) - g_{aFH}^{R*} m_am^2_G C_1   \right], 
	\crn 	\overline{a}^{FHG}_r &= \sum_{F} -\frac{g_{ZGH}  g^{L}_{bFG}}{m_G^2} \left[   g_{aFH}^{R*} m_b\left(m^2_G C_2 +C_{00}\right)\right], 
	\crn 	\overline{b}^{FHG}_l&= \sum_{F} -\frac{g_{ZGH}  g^{L}_{bFG}}{m_G^2} \left[ g^{R*}_{aFH}\left(m_F^2 X_0 +m_b^2X_2- 2m^2_G C_2 \right) -g^{L*}_{aFH}m_am_FX_1\right], 
	\crn	\overline{b}^{FHG}_r &= \sum_{F} -\frac{g_{ZGH}  g^{L}_{bFG} m_b}{m_G^2} \left[ g^{R*}_{aFH}m_aX_1 - g^{L*}_{aFH}m_{F} X_{02} \right],
\end{align}
where arguments for PV-funtions are $(m_{F}^2, m_{H}^2, m_G^2)$. 

The one-loop contributions from the scalar exchanges are well known previously \cite{Jurciukonis:2021izn, Hong:2023rhg}, and we list them here using the general Feynman rules listed in Table \ref{tab_FeynruleZeba}. 
Form factors corresponding to diagrams 4 are 
\begin{align}
	\label{eq_ab4LR}	
	\bar{a}^{FH_kH_l}_{l}&=-\sum_{F}2g_{ZH_kH_l} g^{L*}_{aFH_k} g^{L}_{bFH_l} C_{00},
	\crn  \bar{a}^{FH_kH_l}_{r}& =-\sum_{F}2g_{ZH_kH_l} g^{R*}_{aFH} g^{R}_{bFH} C_{00},
	\crn  \bar{b}^{FH_kH_l}_{l}& =-\sum_{F}2g_{ZH_kH_l} \left[   m_a  g^{L*}_{aFH_k} g^{L}_{bFH_l} X_1  +  m_b  g^{R*}_{aFH_k} g^{R}_{bFH_l}X_2  -m_{F} g^{R*}_{aFH_k} g^{L}_{bFH_l} X_0  \right] ,
	\crn  \bar{b}^{FH_kH_l}_{r}& = -\sum_{F}2g_{ZH_kH_l} \left[   m_a  g^{R*}_{aFH_k} g^{R}_{bFH_l} X_1  +  m_b  g^{L*}_{aFH_k} g^{L}_{bFH_l}X_2  -m_{F} g^{L*}_{aFH_k} g^{R}_{bFH_l} X_0  \right],
\end{align}
where arguments for PV-funtions are $(m_{F}^2, m_{H_k}^2, m_{H_l}^2)$. 

Forms factors corresponding to diagram 6 are
\begin{align}
	\label{eq_ab6LR}	
 \bar{a}^{HFF}_{l}=& -\sum_{F_1,F_2} \left\{  g^{L}_{ZF_{12}}\left[ g^{LL}_{ab}  m_{F_1} m_{F_2}C_0 + g^{RL}_{ab}  m_{a} m_{F_2}(C_0+C_1) \frac{}{}
\right.\right.\crn&\left.\left.\qquad \qquad \qquad
+  g^{LR}_{ab}  m_{b} m_{F_1}(C_0+C_2) +  g^{RR}_{ab}  m_{a} m_bX_0 \frac{}{}\right]   
\right.\crn  &\left. \qquad \quad +\frac{}{} g^{R}_{ZF_{12}} \left[ -g^{LL}_{ab} \left( 2C_{00} +m_a^2 X_1 +m_b^2X_2 -m_Z^2 C_{12}\right) \frac{}{}
\right.\right.\crn&\left.\left.\qquad \qquad \qquad \; \frac{}{}-m_am_{F_1}g^{RL}C_1 -m_bm_{F_2}g^{LR}C_2 \right]\right\}
\crn  \bar{a}^{HFF}_{r}=& -\sum_{F_1,F_2} \left\{  g^{L}_{ZF_{12}} \left[ -g^{RR}_{ab} \left( 2C_{00} +m_a^2 X_1 +m_b^2X_2 -m_Z^2 C_{12}\right) \frac{}{}
\right.\right.\crn&\left.\left.\qquad \qquad \qquad\; -g^{LR} m_am_{F_1}C_1 -g^{RL}m_bm_{F_2}C_2 \frac{}{}\right]
\right.\crn  &\left. \qquad \quad +
\frac{}{} g^{R}_{ZF_{12}}\left[ g^{RR}_{ab}  m_{F_1} m_{F_2}C_0 + g^{LR}_{ab}  m_{a} m_{F_2}(C_0+C_1) \frac{}{}
\right.\right.\crn&\left.\left.\qquad \qquad \qquad \frac{}{}+  g^{RL}_{ab}  m_{b} m_{F_1}(C_0+C_2) +  g^{LL}_{ab}  m_{a} m_bX_0\right] 	\right\},
\crn \bar{b}^{HFF}_{l}=& -\sum_{F_1,F_2}2 \left[\frac{}{} g^{L}_{ZF_{12}}  \left( g^{RL }_{ab} m_{F_2}C_2 +  g^{RR }_{ab}  m_{b} X_2\right) 
	%
+g^{R}_{ZF_{12}}  \left(  g^{RL }_{ab} m_{F_1}C_1 +  g^{LL }_{ab}  m_{a} X_1\right) 
	\right]	,
	\crn\bar{b}^{HFF}_{r} =&- \sum_{F_1,F_2}2 \left[\frac{}{} g^{L}_{ZF_{12}}   \left(  g^{LR }_{ab} m_{F_1}C_1 +  g^{R R}_{ab}  m_{a} X_1\right)
	%
	 +g^{R}_{ZF_{12}} \left( g^{LR }_{ab} m_{F_2}C_2 +  g^{LL }_{ab}  m_{b} X_2\right) 
	\right], 
\end{align}
 where $g^{XY}_{ab}\equiv g^{X*}_{aF_1H} g^{Y }_{bF_2H}$ with $X,Y=L,R$, and   arguments for PV-funtions are $(m_H^2,m^2_{F_1},m^2_{F_2})$. The form factors in Eq. \eqref{eq_ab6LR} are also applicable for Majorana fermions, for example  $ g^{L}_{ZF_{12}} \equiv G_{ij}$ and $ g^{R}_{ZF_{12}} \equiv -G_{ji}$ defined in Eq. \eqref{eq_LintM}. 
 
 The sum of two diagrams 9 and 10 gives 
\begin{align}
	\label{eq_a910LR}
	 \bar{a}^{FH}_{l}&= \sum_F- \frac{ g^{L}_{Ze}}{m_a^2 -m_b^2} \left[  m_{F} \left( m_a g^{RL}_{ab}  + m_b g^{LR}_{ab}   \right) \left(B^{(1)}_0 -B^{(2)}_0\right) 
	\right. \crn& \left. \qquad  \qquad \qquad \qquad -m_am_b g^{RR}_{ab} \left( B^{(1)}_1-B^{(2)}_1\right)   - g^{LL}_{ab}   \left(m_a^2B^{(1)}_1 -m_b^2 B^{(2)}_1 \right) 
	\right],
	\crn \bar{a}^{FH}_{r}&= \sum_F -\frac{ g^{R}_{Ze}}{m_a^2 -m_b^2} \left[  m_{F} \left( m_a g^{LR}_{ab}  + m_b g^{RL}_{ab}   \right) \left(B^{(1)}_0 -B^{(2)}_0 \right) 
	\right. \crn& \left. \qquad \; \qquad \qquad \quad -m_am_b g^{LL}_{ab} \left( B^{(1)}_1 -B^{(2)}_1\right)   - g^{RR}_{ab}   \left( m_a^2B^{(1)}_1 -m_b^2 B^{(2)}_1\right) 
	\right],
\end{align}
where $g^{XY}_{ab}=g^{X*}_{aFH} g^{Y}_{bFH} $ with $X,Y=L,R$, and  $B^{(k)}_{0,1}=B_{0,1}(p_k^2;m_F^2,m_H^2)$. 
\subsection{Particular forms for the 331$\beta$ISS$(K,K)$ model}
 From the general analytical formulas listed in Eq. \eqref{eq_al1u} to Eq. \eqref{eq_a910LR},  the two particular contributions in our work that  $G=W,V$, we compute them based on the following replacements. In particular,  for $G=W$ exchange and $F=n_i$,  we compute them based on the  replacements
\begin{align}
\left\{ F,F_1,F_2,G\right\}	=& \{ n_i,n_i,n_j,W\}, \label{eq_ZWnij} 
	\end{align}
and the general vertex factors given in Table \ref{tab_FeynruleZeba} are matched to the Feynman rules of $W^\pm$, namely, 
\begin{align}
	\left\{g^{L}_{aFG}, g_{ZGG}, g^L_{ZF_{12}}, g^R_{ZF_{12}} \right\}	=& 	\left\{\frac{g}{\sqrt{2}} (U^{\nu\dagger}_0U^e_L)_{ia}, g_{ZW^{+}W^{-}}, G_{ij},-G_{ji} \right\}, \label{eq_gWnij} 
\end{align}
which are shown in Tables \ref{t:hl331iSS} and \ref{table_HGcoupling}; and Eq. \eqref{eq:Gij}. The  formulas of $g^{L(R)}_{Ze}$ are  shown in the second line of Table \ref{tab_Bffp}, consistent with the SM results in the limit $\theta=0$, namely, 
$$  g_{ZW^{+}W^{-}}=t_W^{-1},\; g^L_{Zee}=\frac{-1 +2 s_W^2}{2c_Ws_W} ,\; g^R_{Zee}=t_W. $$
Similarly, the one-loop contributions of $G=V$ exchange and $F=E_a$ are computed easily.

\section{ \label{app:heba} One-loop contributions  to $h\to e_b e_a$ in the 331ISS$(K,K)$ model}
First, we will present here general form factors for all one-loop contributions. The one-loop contributions of a gauge boson $G$ corresponding to diagrams 1 and the sum of  two diagrams 7 and 8 in Fig. \ref{fig_LFVh} are 
\begin{align}
	\Delta^{FGG}_{L} &= -\frac{1}{16\pi^2 } \sum_{F} g^{LL}_{ab} g_{hGG}
	\left\{  -m_{F}^2\left( B^{(1)}_0 + B^{(2)}_0 +B^{(1)}_1\right)  -m_b^2 B^{(2)}_1  + \left(2m_G^2+m^2_{h}\right) m_{F}^2 C_0
	 \right.\crn & \left. \hspace{4.2cm}+ \left[ m_{F}^2 \left( 2m_G^2 +m_{h}^2  \right)+ 2m_G^2\left(2m_G^2 +m_a^2 -m_b^2\right)  \right] C_1 \right. 
	\crn &\left. \hspace{4.2cm}+  
	\left[2m_G^2\left(m_a^2-m^2_{h}\right)+ m_b^2 m^2_{h}\right]C_2\frac{}{}\right\},\crn
\Delta^{FGG}_{R} &= -\frac{1}{16\pi^2 } \sum_{F} g^{LL}_{ab} g_{hGG} 
	\left\{ - m_{F}^2\left( B^{(1)}_0 + B^{(2)}_0 +B^{(2)}_1\right)  -m_a^2 B^{(1)}_1  + \left(2m_G^2+m^2_{h}\right) m_{F}^2 C_0 
	\right.\crn & \left. \hspace{4.2cm}+  \left[ m_{F}^2 \left( 2m_G^2 +m_{h}^2  \right)+ 2m_G^2\left(2m_G^2 -m_a^2 +m_b^2\right)  \right] C_2 \right. 
	\crn &\left. \hspace{4.2cm} +  
	\left[2m_G^2\left(m_b^2-m^2_{h}\right)+ m_a^2 m^2_{h}\right]C_1\frac{}{}\right\},\crn
	\Delta^{FG}_{L}= &\sum_{F} \frac{g  \delta_{hee} g^{LL} m_am_b^2}{(32\pi^2) m_W m_G^2  (m_a^2 -m_b^2) } 
	\crn & \times \left[  2 m_{F}^2\left(B^{(2)}_0-B^{(1)}_0\right) +  \left(2 m_G^2 + m_{F}^2\right) \left(B^{(2)}_1- B^{(1)}_1 \right) + m_b^2 B^{(2)}_1 - m_a^2 B^{(1)}_1  \right]  ,  
	\crn
	\Delta^{FG}_{R} &=\frac{m_a}{m_b}\Delta^{FG}_{L}, 
\end{align}
where $g^{LL} \equiv g^{L*}_{aFG} g^{L}_{bFG}$, and arguments for PV-funtions are $(m_a^2,m_h^2, m_b^2;m^2_{F},m_G^2, m^2_{G})$. 

Diagrams 5 in Fig. \ref{fig_LFVh}: 
\begin{align}
	\Delta^{GFF}_{L}=& \sum_{F_1,F_2} \frac{m_a g^{LL}}{16\pi^2 m_G^2}
	\left\{g^L_{hF_{12}} m_{F_1} \left[\left( B^{(1)}_1 + (2m_G^2+m_{F_2}^2 -m_b^2)\right) C_1 \right] \right.\crn
	& \left. \hspace{2.5cm}+ g^R_{hF_{12}} m_{F_2} \left[ -B^{(12)}_0 +m_G^2C_0 + \left(2m_G^2+ m_{F_1}^2-m_a^2\right)C_1\right]
	\right\}, \crn
	%
	\Delta^{GFF}_{R}=& \sum_{F_1,F_2} \frac{m_b g^{LL}}{16\pi^2 m_G^2}
	\left\{g^L_{hF_{12}} m_{F_1}  \left[ -B^{(12)}_0 +m_G^2C_0 + \left(2m_G^2+m_{F_2}^2-m_b^2\right)C_2\right]\right.\crn
	& \left. \hspace{2.5cm} +   g^R_{hF_{12}} m_{F_2} \left[\left(B^{(2)}_1 + (2m_G^2+m_{F_1}^2 -m_a^2)\right) C_2 \right]\right\}, \nn 
\end{align}
where $g^{LL}\equiv g^{L*}_{aF_1G} g^{L}_{bF_2G}$, and arguments for PV-funtions are $(m_G^2,m^2_{F_1}, m^2_{F_2})$. 

The one-loop contributions from scalar exchange in diagrams 4, 6, and  the sum of the two diagrams 9 and 10 are denoted as $\Delta^{FH_1H_2}_{L(R)}$,  $\Delta^{H F_1F_2}_{L(R)}$, and $\Delta^{ FH}_{L(R)}$. The particular formulas of $\Delta^{FH_1H_2}_{L(R)}$ and  $\Delta^{ FH}_{L(R)}$are 
\begin{align}
\Delta^{FH_1H_2}_{L}& =\frac{\lambda_{hH_1H_2}}{16\pi^2}\sum_{F}   \left[g^{RL}m_{F}C_0  -g^{LL}m_{a}C_1 - g^{RR} m_{b}C_2 \right], \label{eq:deFH12L}\\
	\Delta^{FH_1H_2}_{R} &= \frac{\lambda_{hH_1H_2}}{16\pi^2} \sum_{F} \left[ g^{LR}m_{F}C_0  - g^{RR}m_{a} C_1 - g^{LL} m_{b}C_2 \right],	 \label{eq:FH12R} \\
%
\Delta^{FH}_{L}  =& \frac{g\delta_{hee}}{32\pi^2m_W \left(m_a^2-m_b^2\right)} 
\sum_{F} \left[ m_am_b m_{F} g^{RL}  \left(B^{(1)}_0- B^{(2)}_0\right) + m_{F}
g^{LR} \left(m^2_bB^{(1)}_0-m^2_aB^{(2)}_0\right)
\right.\crn &\left. \hspace{4.5cm} - m_{a}m_b \left( g^{RR} m_a + g^{LL} m_b\right)\left(B^{(1)}_1- B^{(2)}_1\right)\right], \label{eq:deFHL}\\
\Delta^{F H}_{R}  =& \frac{g\delta_{hee}}{32 \pi^2m_W\left(m_a^2-m_b^2\right)}  \sum_{F} \left[ m_am_bm_{F} g^{LR}\left(B^{(1)}_0-B^{(2)}_0\right)+ m_{F}
g^{RL} \left(m^2_bB^{(1)}_0-m^2_aB^{(2)}_0\right) 
\right.\crn&\left.  \hspace{4.5cm}- m_{a}m_b \left( g^{LL}m_a + g^{RR} m_b\right)\left( B^{(1)}_1 - B^{(2)}_1\right)\right], \label{eq:deFHR}
\end{align}
where $ g^{XY}_{ab}\equiv g^{X*}_{aFH_1} g^{Y}_{bFH_2}$ for $X,Y=L,R$, and arguments for PV-funtions are $(m^2_{F},m^2_{H_1}, m^2_{H_2})$.  

 The particular formulas of $\Delta^{H F_1F_2}_{L(R)}$ are  
\begin{align}
	\Delta^{HFF}_{L} &=    \frac{1 }{16\pi^2}\sum_{F_1,F_2} \left\{ g^L_{hF_{12}}\left[ g^{RL} m_{F_1}m_{F_2}C_0 +g^{LR} m_{a}m_{b}X_0 \frac{}{}\right.\right.\crn
	&\left.\left. \hspace{3.5cm}
	+g^{LL} m_{a}m_{F_2}(C_0 +C_1)+  g^{RR} m_{b} m_{F_1}(C_0+C_2) \frac{}{} \right]  \right. \crn
	& \left. \hspace{2.2cm}\frac{}{}  + g^R_{hF_{12}} \left[ g^{RL} \left(B^{(12)}_0+m_{H}^2C_0 +m_a^2 C_1 +m_b^2C_2\right)\right.\right.\crn
	&\left.\left. \hspace{3.6cm}\frac{}{}+ g^{LL} m_am_{F_1}C_1 +g^{RR} m_bm_{F_2}C_2  \right]\right\}, \label{eq:deHFFL}\\
	\Delta^{HFF}_{R} &=      \frac{1 }{16\pi^2}\sum_{F_1,F_2} \left\{ g^L_{hF_{12}}\left[ g^{LR} \left(B^{(12)}_0+m_{H}^2C_0 +m_a^2 C_1 +m_b^2C_2\right)\right.\right.\crn
	&\left.\left. \frac{}{}\hspace{3.5cm}  +g^{RR}m_am_{F_1} C_1 +g^{LL} m_bm_{F_2}C_2 \right]\right. \crn
	&\left.\hspace{2.4cm} + g^R_{hF_{12}} \left[ g^{LR} m_{F_1}m_{F_2}C_0 + g^{RL} m_{a}m_{b}X_0  \frac{}{}\right.\right.\crn
	&\left.\left.\frac{}{} \hspace{3.5cm}+  g^{RR} m_{a}m_{F_2}(C_0 +C_1) + g^{LL} m_{b}m_{F_2} (C_0+C_2) \right]\right\},
\end{align}
where  $ g^{XY}\equiv g^{X*}_{aF_1H} g^{Y}_{bF_2H}$ for $X,Y=L,R$, and  The arguments of  PV functions are  $(p_1^2,m_h^2,p_2^2;m_{H}^2,m_{F_1}^2, m_{F_2}^2)$. 

One-loop contributions from two diagrams 2 and 3 in Fig. \ref{fig_LFVh} containing both Higgs and gauge bosons  were introduced previously \cite{Hue:2015fbb, Hong:2022xjg}, but they vanish  in our numerical investigation because of the suppressed factors $s_{\theta}$ and $s_{\delta}$, which are fixed as zeros. 

The  one-loop contributions given in Eq. \eqref{deLR} are derived by the following transformations:
\begin{align*}
	\Delta^{(1)W}_{L(R)} =& \Delta^{nWW}_{L(R)}=\sum_{i=1}^{3+2K}\Delta^{FGG}_{L(R)} \left[F\to n_i,G\to W\right],  \\
\Delta^{(5)W}_{L(R)} =& \Delta^{Wnn}_{L(R)}=\sum_{i,j=1}^{3+2K}\Delta^{GF_1F_2}_{L(R)} \left[F_1\to n_i, F_2\to n_j,G\to W\right],  \\
	\Delta^{(7+8)W}_{L(R)} =& \Delta^{nW}_{L(R)}=\sum_{i=1}^{3+2K}\Delta^{FG}_{L(R)} \left[F\to n_i,G\to W\right], \\
\Delta^{(4)H}_{L(R)} =& \sum_{k,l=1}^2\Delta^{nH^\pm_kH^\pm_l}_{L(R)}=\sum_{k,l=1}^2\sum_{i=1}^{3+2K}\Delta^{FH_1H_2}_{L(R)} \left[F\to n_i,H_1\to H^{\pm}_k, H_2\to H^{\pm}_l\right],  \\
\Delta^{(6)H}_{L(R)} =& \sum_{k}^2\Delta^{H^\pm_knn}_{L(R)}=\sum_{k}^2\sum_{i,j=1}^{3+2K}\Delta^{HF_1,F_2}_{L(R)} \left[H\to H^{\pm}_k, F_1\to n_i, F_2\to n_j,\right],  \\
\Delta^{(9+10)H}_{L(R)} =& \sum_{k,l=1}^2\Delta^{nH^\pm_k}_{L(R)}=\sum_{k=1}^2\sum_{i=1}^{3+2K}\Delta^{FH}_{L(R)} \left[F\to n_i,H\to H^{\pm}_k\right]. 
\end{align*}

\end{document}